\documentclass[twocolumn,showpacs,superscriptaddress,preprintnumbers,amsmath,amssymb,longbibliography,prl]{revtex4-1}
\usepackage{graphicx}
\usepackage{dcolumn}
\usepackage{bm}
\usepackage{amsmath}
\usepackage{color}

\begin{document}

\title{Critical scaling for yield is independent from distance to isostaticity}
\author{Jacob D. Thompson and Abram H. Clark}
\affiliation{Department of Physics, Naval Postgraduate School, Monterey, California 93943, USA}

\begin{abstract}

Using discrete element simulations, we demonstrate that critical behavior for yielding in soft disk and sphere packings is independent of distance to isostaticity over a wide range of dimensionless pressures. Jammed states are explored via quasistatic shear at fixed pressure, and the statistics of the dimensionless shear stress $\mu$ of these states obey a scaling description with diverging length scale $\xi \propto |\mu-\mu_c|^{-\nu}$. The critical scaling functions and values of the scaling exponents are nearly independent of distance to isostaticity despite the large range of pressures studied. Our results demonstrate that yielding of jammed systems represents a distinct nonequilibrium critical transition from the isostatic critical transition which has been demonstrated by previous studies. Our results may also be useful in deriving nonlocal rheological descriptions of granular materials, foams, emulsions, and other soft particulate materials. 

\end{abstract}

\date{\today}

\maketitle

Granular materials, dense suspensions, foams, and emulsions can form amorphous jammed states~\cite{jaeger1996,ohern2003,donev2004,vanhecke2009,liu2010jamming}. Jammed systems can \textit{yield} when subjected to a sufficient shear stress $\tau$ (this is sometimes called ``unjamming by shear''). When $\mu\equiv \tau/p$, where $p$ is the system pressure, exceeds a critical value $\mu_c$, jammed states become inaccessible and flow persists indefinitely~\cite{drucker1952,gardiner1998,toiya2004,dacruz2005,xu2006,jop2006,peyneau08,peyneau08-2}. Predicting the mechanical response of jammed states can be difficult since it can involve plastic rearrangement events that cooperate over large distances. For example, rheological models of these materials that include nonlocal cooperative effects can successfully reproduce steady-state flows from experiments and particle-based simulations~\cite{goyon2008,masselon2008,bocquet2009,jop2012,kamrin2012,henann2013,bouzid2013,bouzid2015}.

Prior studies on soft sphere packings, which are commonly used to model these materials, have framed long-range cooperative behavior in terms of a nonequilibrium critical transition that occurs at the isostatic point, also called ``point J"~\cite{ohern2003,drocco2005multiscaling,ellenbroek2006critical,olsson2007,Nordstrom2010,Tighe2010,olsson2011,paredes2013rheology}. Isostaticity refers to the number of contacts per particle $Z$ being equal to the number required to constrain all degrees of freedom in the system, $Z=Z_{\rm iso}$. This occurs at a given volume fraction $\phi=\phi_c$ in the large-system limit. At isostaticity, $p=0$, but further compression leads to increasing $p$. A cooperative length scale $\xi_J \propto |\phi-\phi_J|^{-\nu_J}$ diverges at the isostatic point, which then controls the mechanical response and leads to Widom-like scaling relations~\cite{goodrich2016scaling} that relate $p$, $(\phi-\phi_J)$, $(Z-Z_{\rm iso})$, and other quantities. $\xi_J$ is large near isostaticity (i.e., small $p$), characterized by an excess of spatially extended, low-energy modes of the system~\cite{silbert2005vibrations,wyart2005effects}. For increasing $p$, $\xi_J$ decreases, leading to more localized modes as well as smaller and more localized particle rearrangements.

In contrast, nonlocal rheological descriptions of jammed materials~\cite{bocquet2009,kamrin2012,henann2013,bouzid2013,bouzid2015,ClarkYielding} often include a diverging cooperative length scale that depends not on packing fraction but on distance to a critical \textit{shear stress}, i.e., $\xi \propto |\mu - \mu_c|^{-\nu}$. These rheological models, including our previous paper~\cite{ClarkYielding}, describe materials that are near $\phi=\phi_J$, so it is not known how the cooperative length scale underlying these models relates to the isostatic critical point. Here we show using numerical simulations that yielding in soft sphere packings is a distinct nonequilibrium critical transition and that it is independent from distance to isostaticity. We quasistatically shear systems of repulsive, bidisperse disks and spheres, holding dimensionless pressure $\tilde{p}$ fixed and measuring $\mu$, which increases during an initial shear regime and then plateaus as stress is released in intermittent slips. The statistics of $\mu$ obey a scaling description with a diverging length scale $\xi \propto |\mu - \mu_c|^{-\nu}$, where $\nu_{\rm ms}\approx 1.8$ during initial shear buildup (in agreement with \cite{ClarkYielding}) and $\nu_{\rm slip}\approx 1.1$ in two dimensions (2D) and $\nu_{\rm slip}\approx 0.8$ in three dimensions (3D) during the intermittent slip regime. The scaling functions and the values of $\nu$ are highly insensitive to the distance from isostaticity set by $\tilde{p}$, which we vary over nearly four orders of magnitude. $\mu_c(\tilde{p})$ is insensitive to $\tilde{p}$ for $\tilde{p} \le 10^{-3}$, but decreases logarithmically for higher $\tilde{p}$. The critical scaling functions we show could be used to derive a particle-scale theory for nonlocal rheological models, including transient behavior, which is not captured by current models.

\begin{figure}
\raggedright
(a) \hspace{37mm} (b) \\
\includegraphics[trim=40mm 10mm 40mm 10mm, clip, width=0.49\columnwidth]{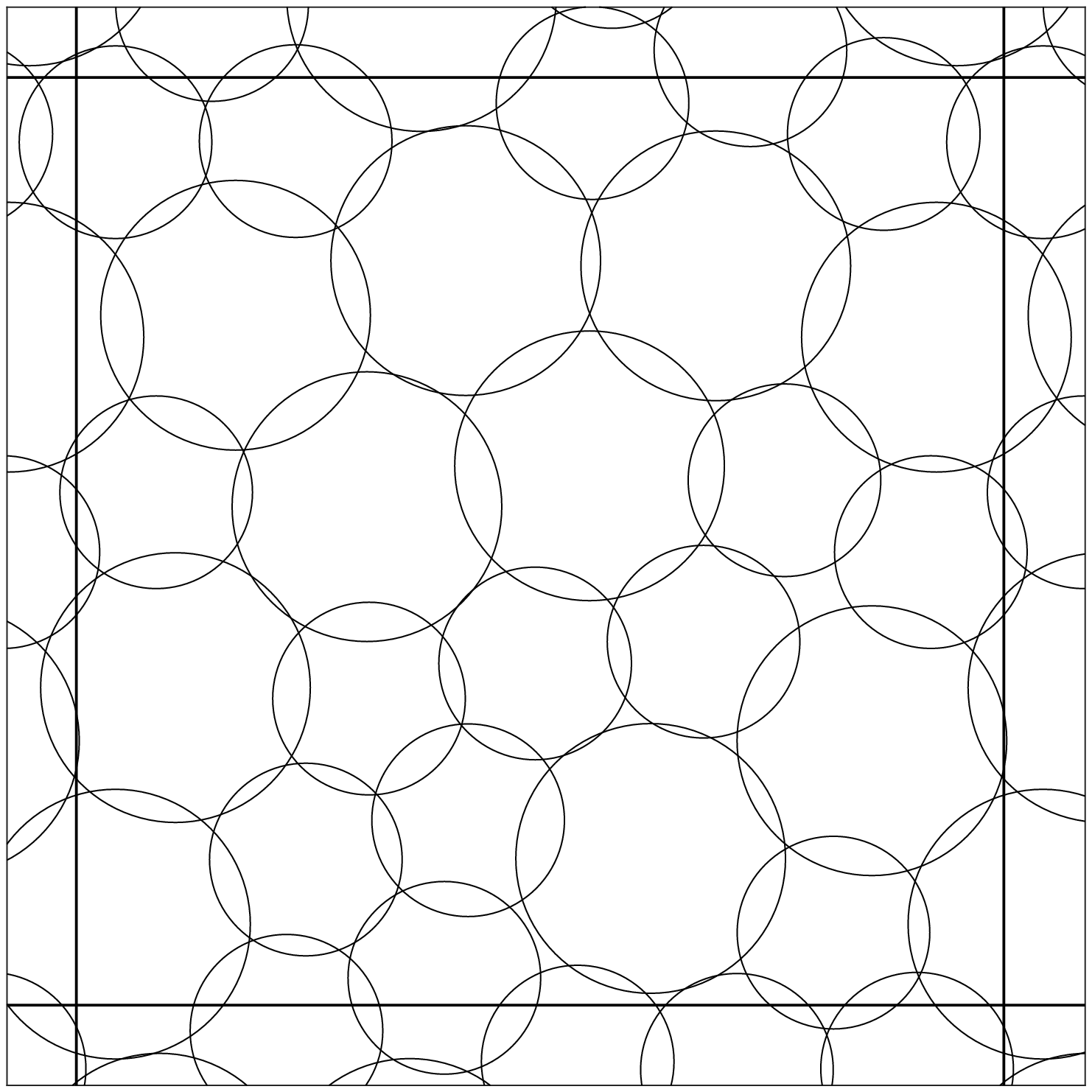}
\includegraphics[trim=40mm 10mm 40mm 10mm, clip, width=0.49\columnwidth]{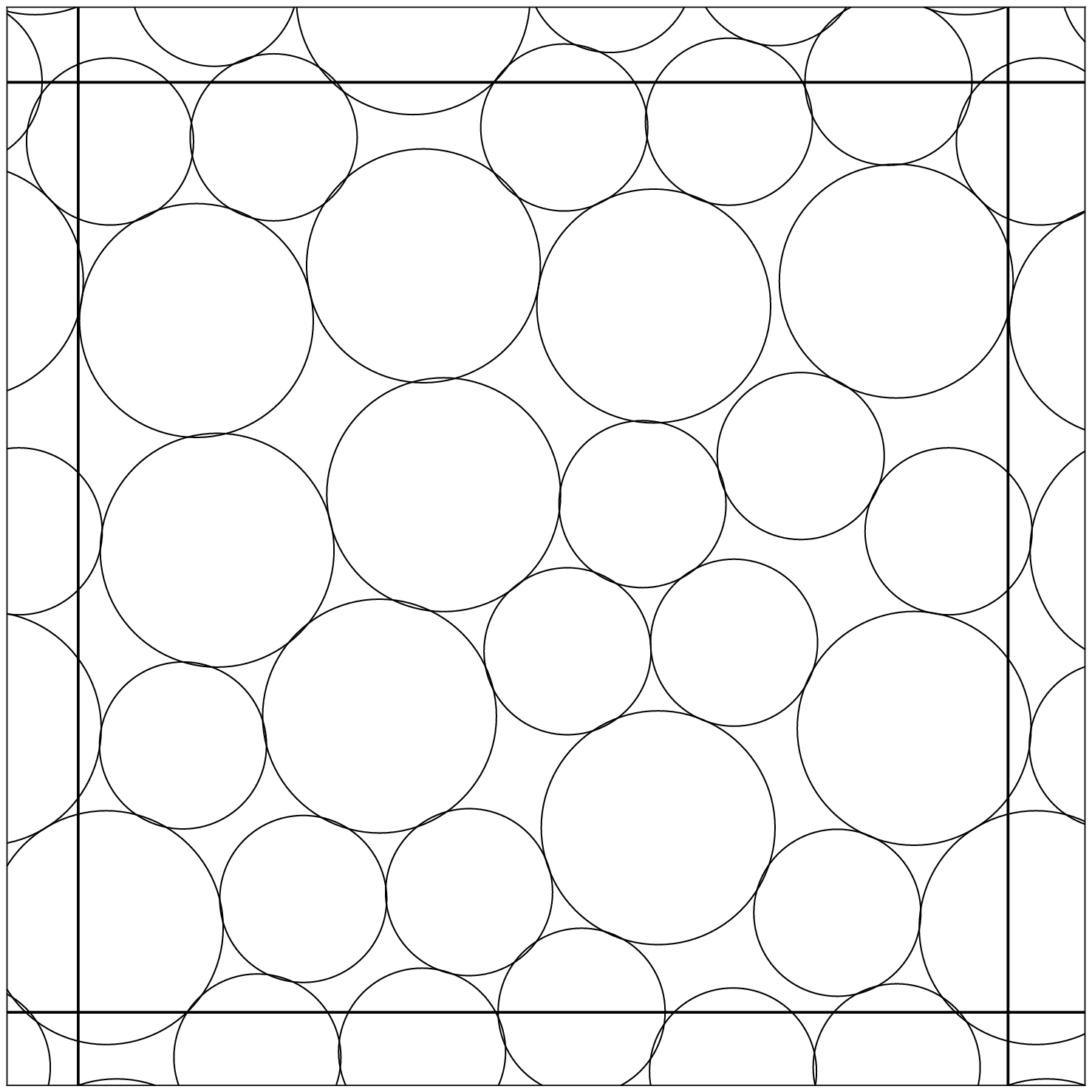} \\
\raggedright
(c)\\
\includegraphics[trim=0mm 0mm 25mm 0mm, clip, width = 1.0\columnwidth]{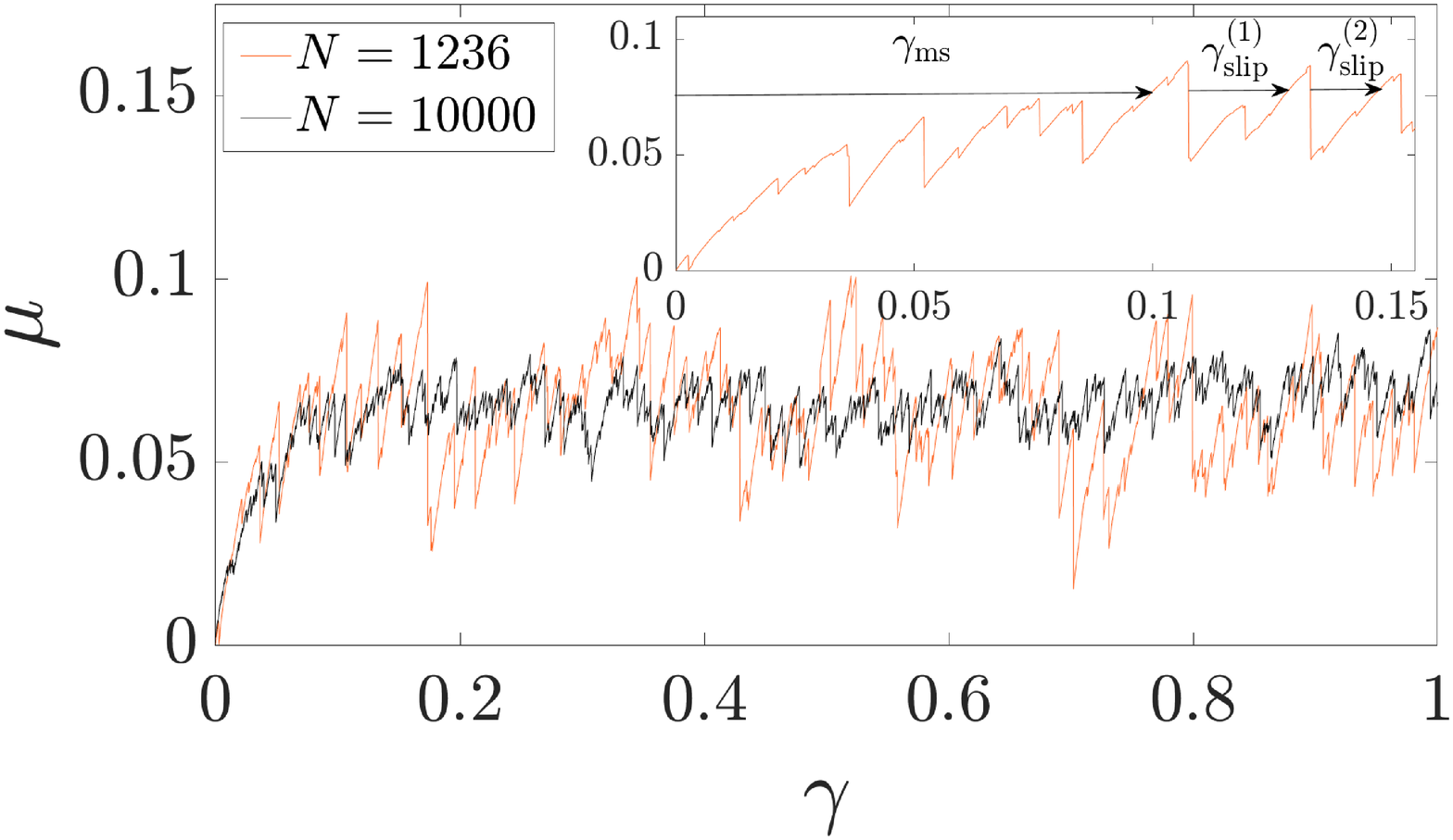}
\caption{(a, b) Illustrative snapshots during Lees-Edwards shear at $\gamma = 0.33$ with dimensionless pressure (a) $\tilde{p} \equiv p/K = 0.2$ (far from isostaticity) and (b) $\tilde{p} = 0.001$ (near isostaticity). (c) Plot of dimensionless shear stress $\mu$ versus $\gamma$ for a single simulation with 1,236 (orange) and 10,000 (black) particles. The inset shows a closeup of $0\le\gamma\le 0.15$. The first arrow indicates the initial shear strain $\gamma_{\rm ms}$ required to find the first state at a particular value of $\mu$ (in the case shown, $\mu \approx 0.077$). Subsequent arrows denote the shear strains $\gamma_{\rm slip}$ between states where the shear stress is less than the value of $\mu$ being considered.}
\label{fig:Setup}
\end{figure}

\textit{Methods.---} We use molecular dynamics simulations to study systems of $N$ bidisperse frictionless disks in 2D and spheres in 3D, with diameter ratio $1.4$ and equal numbers of each size. Systems are prepared at a given pressure $p$ via isotropic compression and then quasistatically sheared. Contacting particles interact via a purely repulsive force $\mathbf{F}_{ij} = K (\delta_{ij}/ |\mathbf{r}_{ij}| - 1)\mathbf{r}_{ij}$, where $\delta_{ij}$ is the average diameter of particles $i$ and $j$, $\mathbf{r}_{ij}$ is the vector displacement between the centers of particles $i$ and $j$. Stresses are quantified by the Cauchy stress tensor,
\begin{equation}
\sigma_{\alpha \lambda} = \frac{1}{V} \sum\limits_{i\neq j} r^{ij}_\alpha F^{ij}_\lambda.
\label{eqn:stress-tensor}
\end{equation}
Here, $\alpha$ and $\lambda$ are Cartesian coordinates, $V$ is the system volume, $r^{ij}_\alpha$ is the $\alpha$-component of the center-to-center separation vector between particles $i$ and $j$, and $F^{ij}_\lambda$ is the $\lambda$-component of the interparticle contact force. The sum over $i$ and $j$ includes all pairs of contacting particles. 

Each simulation is prepared by placing particles randomly throughout the domain and then increasing the particle diameter $D$ in small steps until reaching a target $p = (\sigma_{xx}+\sigma_{yy})/2$. Using Lees-Edwards boundary conditions, we impose affine shear strain in small steps $\Delta \gamma = 10^{-4}$. At each shear step, the shear-periodic boundary is moved by $\Delta \gamma$ and $y \Delta \gamma$ is added to the $x$-position of each particle. We then use molecular dynamics to relax the potential energy using modified velocity Verlet integration, as well as shrink or swell $D$ to maintain a fixed $p$ within $0.5\%$ of the target value. Before shearing or changing the particle diameter, we damp out kinetic energy via a viscous damping force $-B\mathbf{v}$ to each particle, where $\mathbf{v}$ is the absolute velocity of a given particle and $B$ is the damping coefficient. We set $B=5 \sqrt{M p}$, where $M$ is the mass of a large grain. Our results are independent of $B$ in this regime. 

At each strain step, after the system is quenched at the target pressure, we measure the stress tensor elements, as defined in Eq.~\eqref{eqn:stress-tensor}, focusing on $\mu = -\sigma_{xy} / p$, as shown in Fig.~\ref{fig:Setup}(c). We measure $\mu$ from $0\le\gamma \le 3$ in increments of $\Delta \gamma = 10^{-4}$ for a total of 30,001 states per simulation. For each value of $N$ and $\tilde{p}$, we simulate an ensemble of 400 systems.

\begin{figure}
\raggedright
(a) \hspace{37mm} (b) \\
\includegraphics[width=0.49\columnwidth]{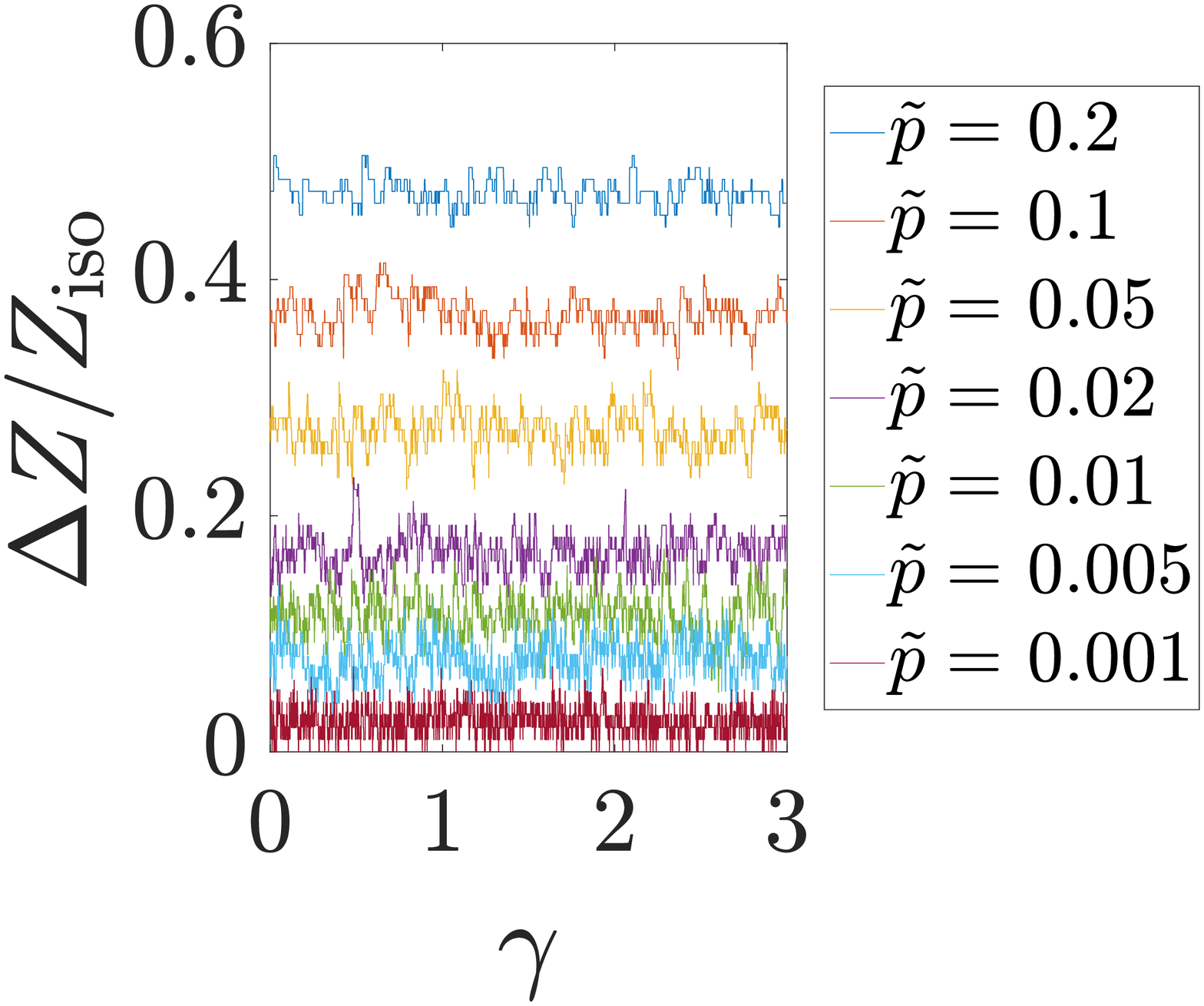}
\includegraphics[width=0.49\columnwidth]{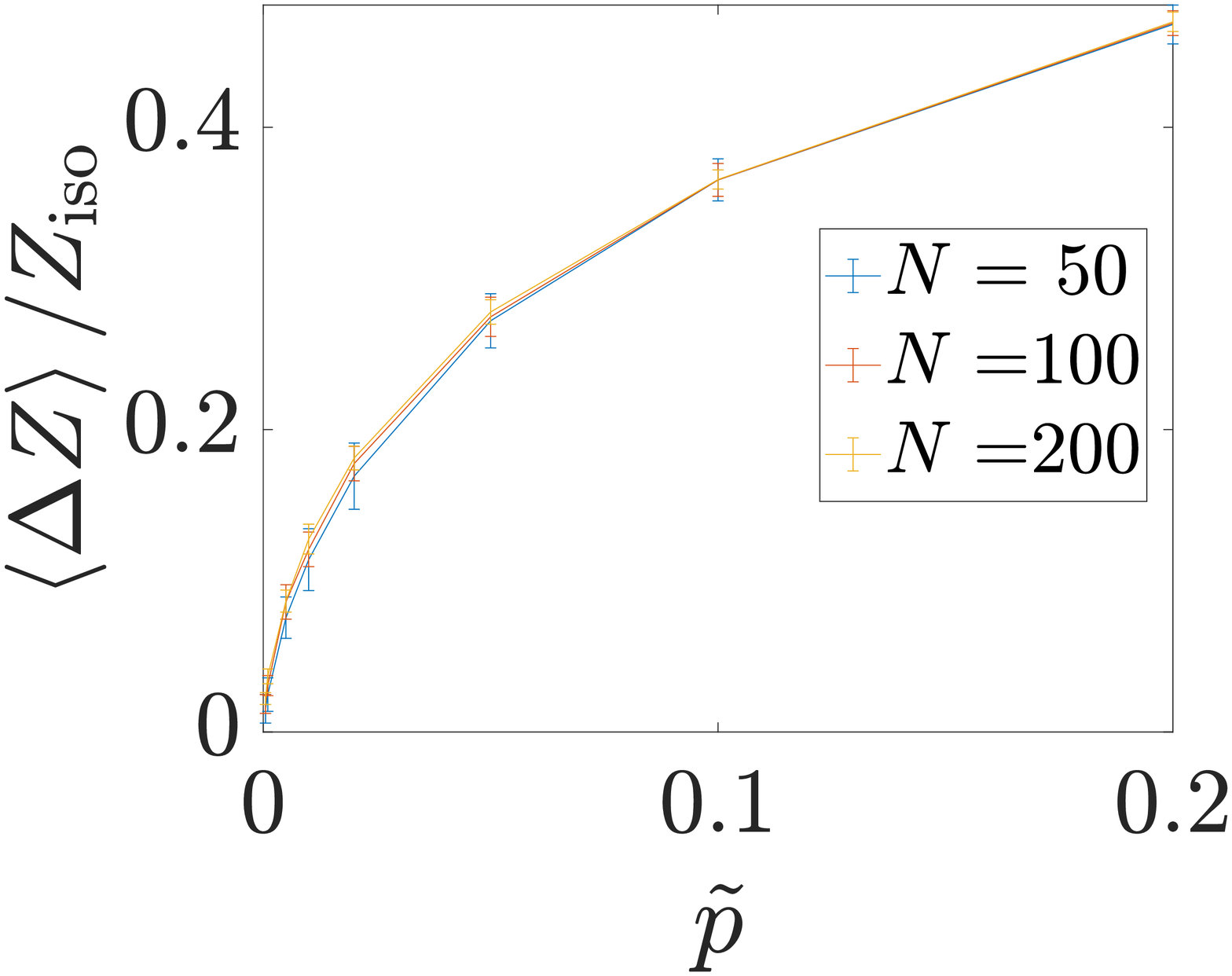}
\caption{(a) Excess contacts $\Delta Z/Z_{\rm iso}$ versus the shear strain $\gamma$ for individual simulations with $N=50$, where $\Delta Z\equiv Z - Z_{\rm iso}$, $Z$ is the coordination number once rattlers have been removed, and $Z_{\rm iso}$ is the coordination number for an isostatic system. (b) Mean $\Delta Z/Z_{\rm iso}$ versus $\tilde{p}$ over 50 simulations for $N=50$, 100 and 200, showing that $\tilde{p}$ gives the fraction of extra contacts, independent of system size.}
\label{fig:Contacts}
\end{figure}

\begin{figure*}[t]
\raggedright
(a) \hspace{55mm} (b) \hspace{55mm} (c) \\
\includegraphics[width=0.68\columnwidth]{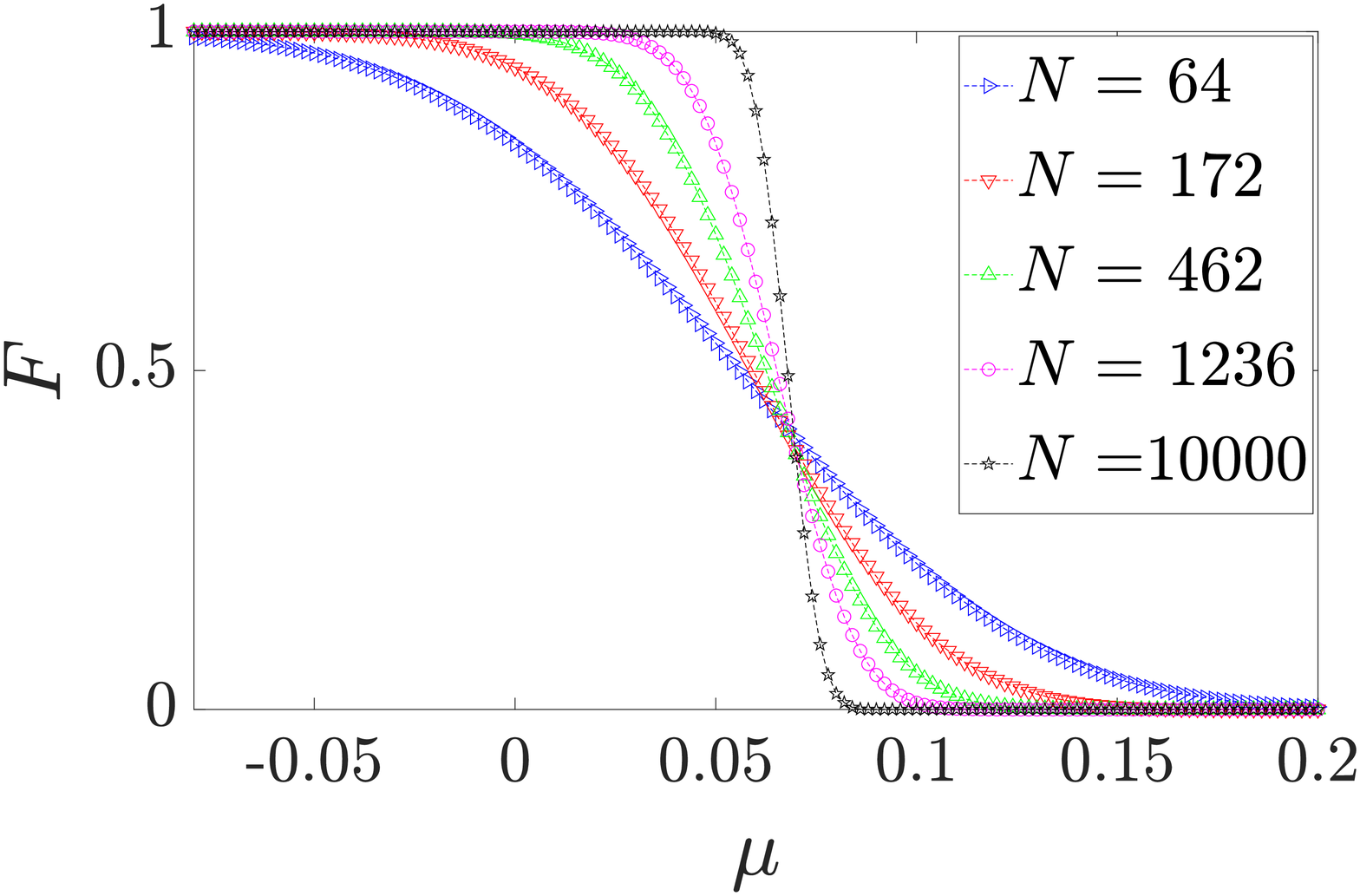}
\includegraphics[width=0.68\columnwidth]{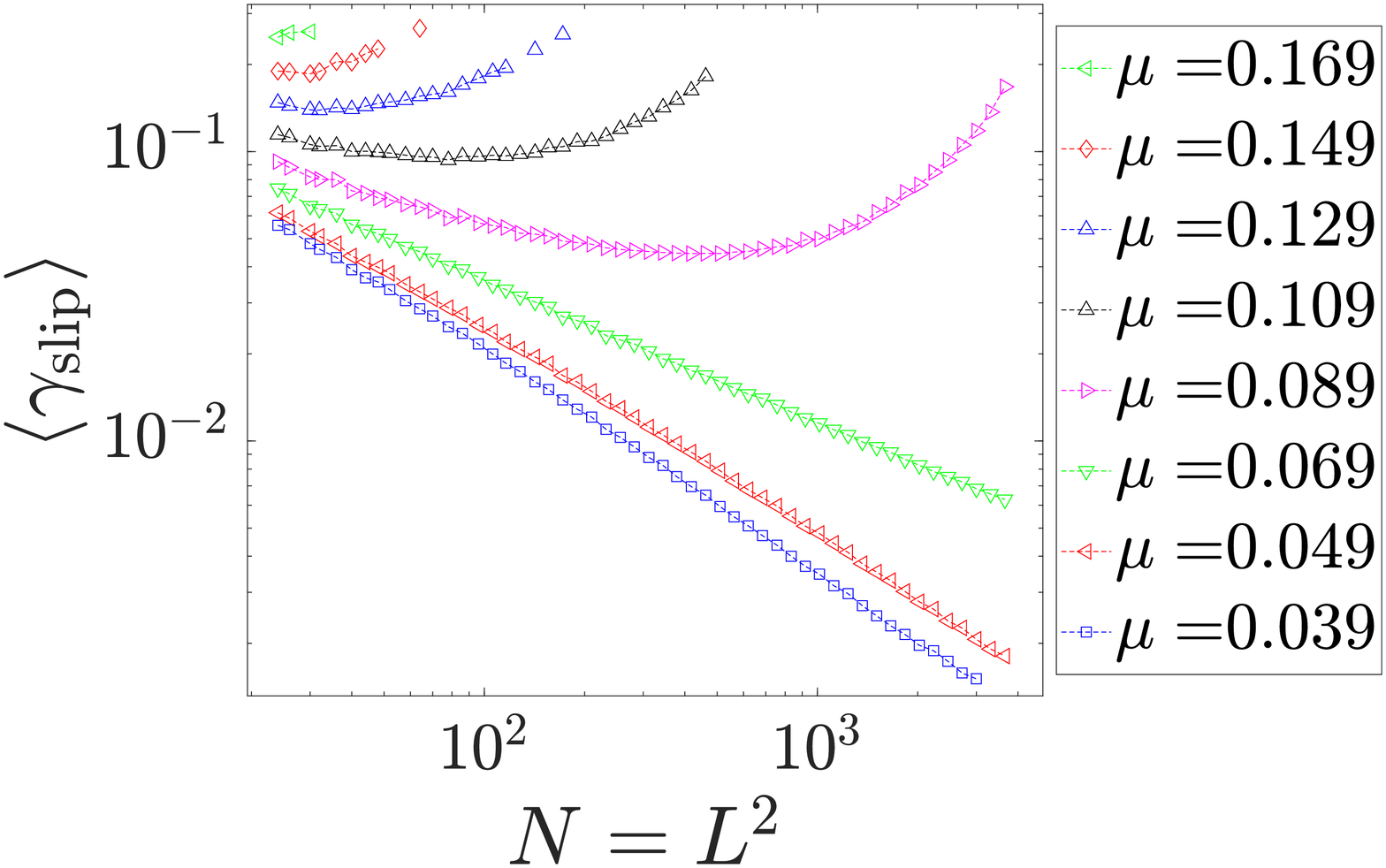}
\includegraphics[width=0.68\columnwidth]{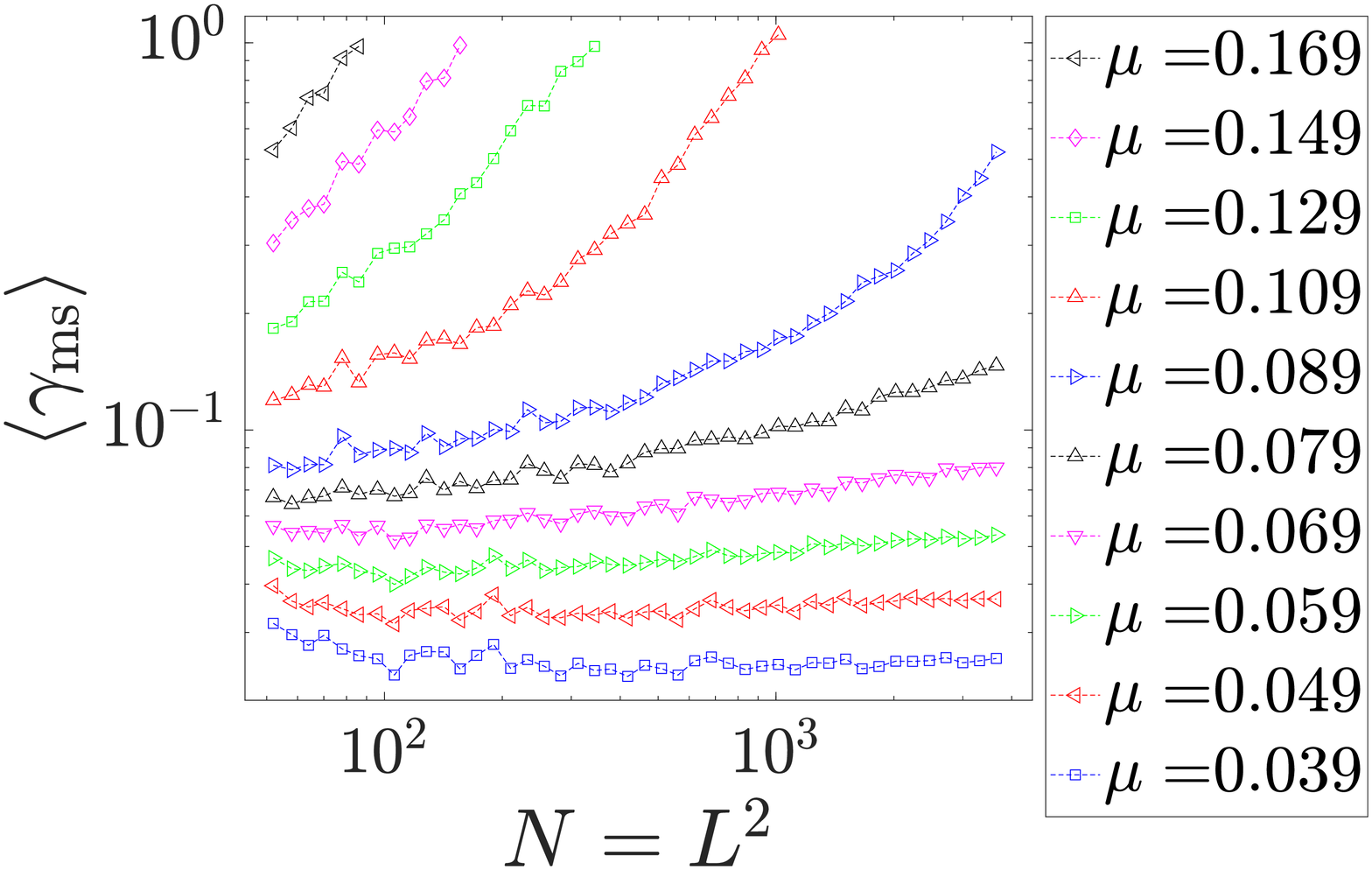} \\
(d) \hspace{55mm} (e) \hspace{55mm} (f) \hspace{55mm} \\
\includegraphics[width=0.68\columnwidth]{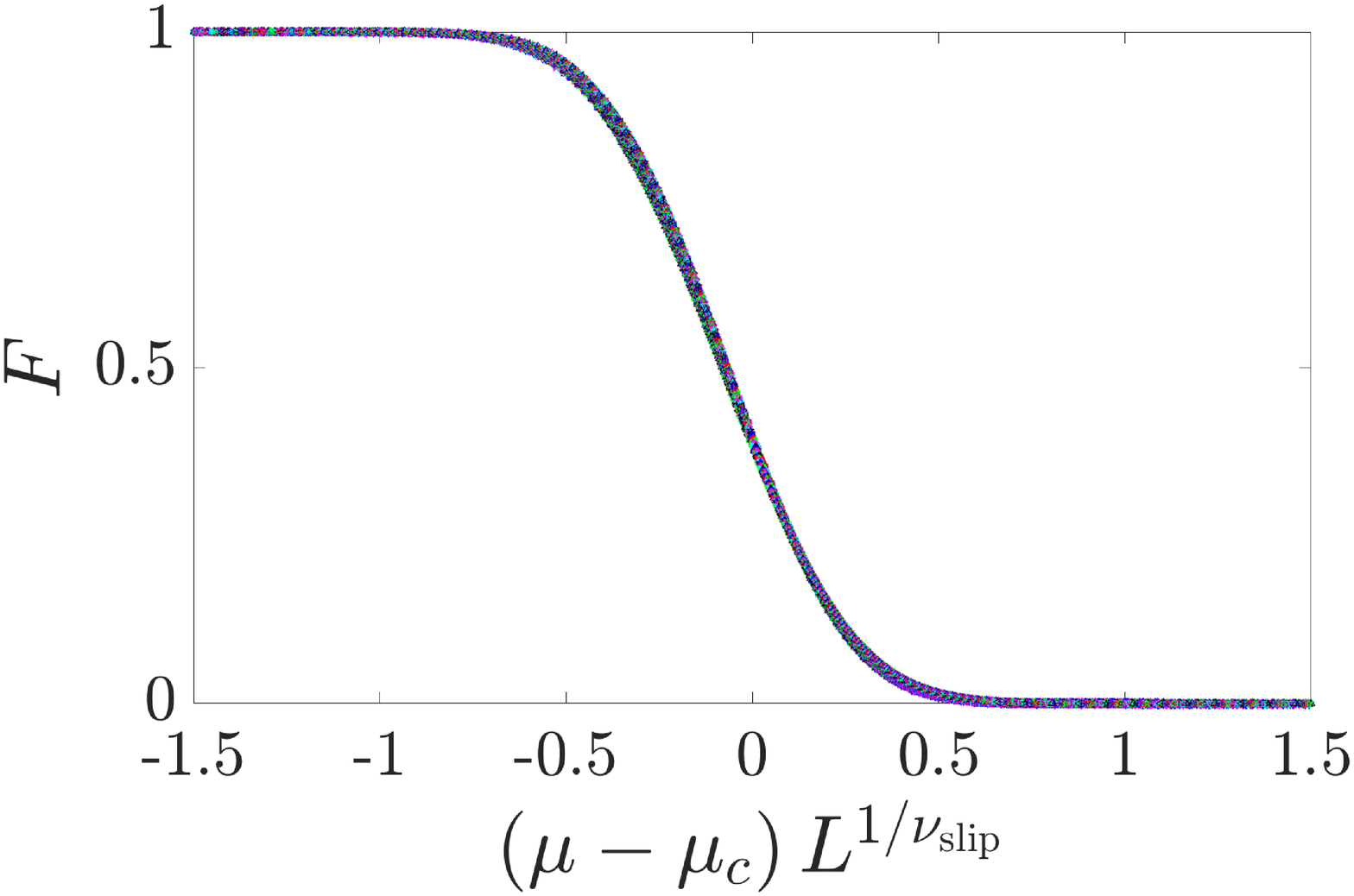}
\includegraphics[width=0.68\columnwidth]{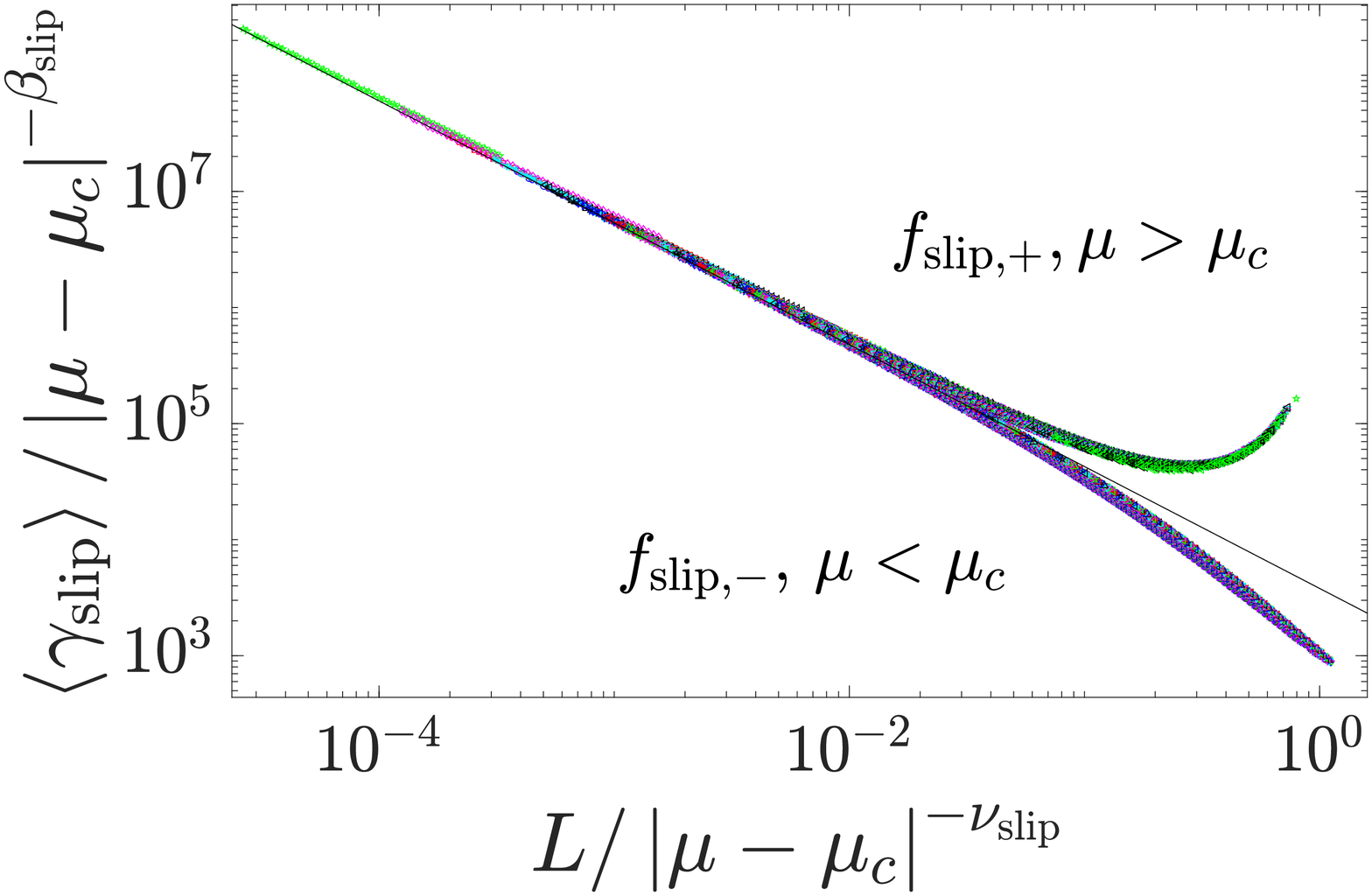}
\includegraphics[width=0.68\columnwidth]{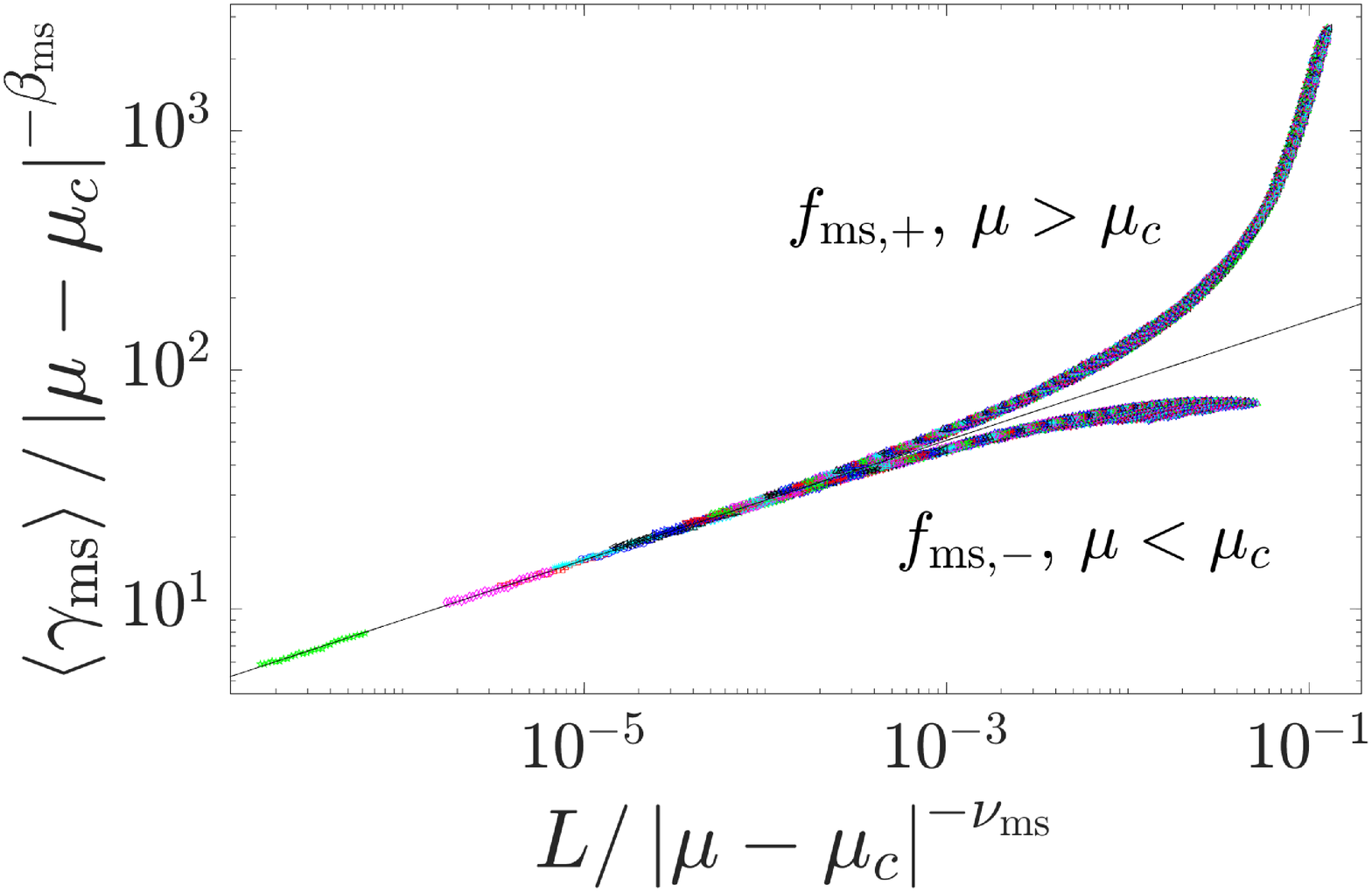}
\caption{Scaling collapses in 2D at high dimensionless pressure $\tilde{p}=0.05$. Fraction of states $F$ above $\mu$, unscaled (a) and scaled (d), with $\mu_c=0.061$ and $\nu_{\rm slip}=1.09$. $\left\langle\gamma_{\rm slip}\right\rangle$ versus $N$, unscaled (b) and scaled (e), for $24\le N\le 3654$, with $\mu_c = 0.061$, $\nu_{\rm slip} = 1.09$, and $\beta_{\rm slip}/\nu_{\rm slip} = -1.05$. (c), (f) Mean strain $\left\langle\gamma_{\rm ms}\right\rangle$, unscaled (c) and scaled (f), to the first mechanically stable (MS) state at dimensionless shear stress $\mu$, for $24\le N\le 3654$, collapsed onto the proposed scaling form with $\mu_c = 0.061$, $\nu_{\rm ms} = 1.8$, and $\beta_{\rm ms}/\nu_{\rm ms} = 0.25$. In all three cases, the scaled data includes all values of $\mu$ and $N$, and the unscaled plots show only selected curves. The data for $F$ is plotted versus $\mu$ with different curves representing different $N$, and $\left\langle\gamma_{\rm slip}\right\rangle$ and $\left\langle\gamma_{\rm ms}\right\rangle$ are plotted versus system size with different curves representing different values of $\mu$.}
\label{fig:Unscaled}
\end{figure*}


We quantify distance above isostaticity by $\tilde{p} = p/K$, which gives an estimate of the relative overlap between particles (i.e., $\tilde{p} = 0.01$ corresponds to particle-particle overlaps of roughly $0.01D$). Figure~\ref{fig:Setup}(a) and (b) show $\tilde{p} = 0.2$ and $\tilde{p}=0.001$, respectively, with $N=24$. Overcompression yields excess contacts such that $\Delta Z / Z_{\rm iso}\sim \sqrt{\tilde{p}}$, where $\Delta Z \equiv (Z - Z_{\rm iso})$, $Z$ is the number of contacts per particle, $Z_{\rm iso}\equiv 2\left(d\left(N-N_r\right)-d+1\right)/\left(N-N_r\right)$ is the isostatic number of contacts, $d$ is the number of spatial dimensions, and $N_r$ is the number of rattlers \cite{vanhecke2009,papanikolaou2013,Shen2014,VanderWerf2018}. Figure~\ref{fig:Contacts}(a) shows $\Delta Z/Z_{\rm iso}$ plotted versus $\gamma$ for $N=50$ and varying $\tilde{p}$. These curves fluctuate around a fixed value but show no trend, indicating that the shearing does not change $Z$ on average. Figure~\ref{fig:Contacts}(b) shows that the average value $\langle \Delta Z \rangle /Z_{\rm iso}$ versus $\tilde{p}$ is similar for $N=50$, $100$, and $200$. Thus, the fraction of excess contacts and thus the distance to isostaticity is set by $\tilde{p}$, nearly independent of system size~\cite{GoodrichFiniteSizeScaling} or the presence of shear deformation~\cite{FavierDeCoulomb2017}.

\textit{Scaling near yield.---} As shown in Fig.~\ref{fig:Setup}(c), $\mu$ increases with $\gamma$ and then plateaus as potential energy is released in intermittent slips~\cite{Miguel2001,Dalton2001,Maloney2006,Daniels2008,dahmen2011,Salerno2012,Bares2017}. This curve represents a series of jammed states that the system passes through while sheared. The fluctuations in $\mu$ decrease with increasing $N$ for all $\tilde{p}$, and we exploit the size scaling in these fluctuations (as in \cite{lin2014}) to demonstrate and quantify diverging spatial correlations near $\mu_c$. Most importantly, we show that this scaling description is nearly independent from the distance to isostaticity. 

To accomplish this, we use finite size scaling on three quantities for each $\mu$ and $N$: (1) the cumulative distribution function $F$ of states above a particular value of $\mu$ during the slip avalanche regime, defined as $\gamma>0.5$ (our results are insensitive to this choice); (2) the shear strain $\gamma_{\rm slip}$ between mechanically stable (MS) states with an internal shear stress of at least $\mu$; and (3) the shear strain $\gamma_{\rm ms}$ required to find the first MS state with an internal shear stress of at least $\mu$. Figure~\ref{fig:Setup}(c) depicts $\gamma_{\rm ms}$ and $\gamma_{\rm slip}$ for a given $\mu(\gamma)$ curve. Figure~\ref{fig:Unscaled}(a)-(c) shows these quantities plotted as functions of $\mu$ or $N$, where ensemble averages are denoted with angle brackets. The data shown in Fig.~\ref{fig:Unscaled} represents only a single value of $\tilde{p}=0.05$ in 2D, but it is typical of all $\tilde{p}$ in both 2D and 3D, as we demonstrate below in Figs.~\ref{fig:K2000} and~\ref{fig:3D}. As $N$ is increased, the fluctuations in $\mu$ decrease, and $F$ approaches a step function, as shown in Fig.~\ref{fig:Unscaled}(a). Thus, MS states vanish sharply at some value $\mu=\mu_c(\tilde{p})$ in the large system limit. Figure~\ref{fig:Unscaled}(b) shows $\langle \gamma_{\rm slip} \rangle$, where we require at least one $\gamma_{\rm slip}$ measurement per simulation. Our results are insensitive to this choice, unless the number of samples becomes very small. For $\mu<\mu_c$, $\langle \gamma_{\rm slip} \rangle$ monotonically decreases with increasing $N$. For $\mu>\mu_c$, $\langle \gamma_{\rm slip} \rangle$ first decreases and then increases with increasing $N$. Finally, $\langle \gamma_{\rm slip} \rangle$ is nearly independent of $N$ for small $\mu$ and increases strongly with $N$ for larger $\mu$.

To collapse these curves, we posit a diverging length scale $\xi \propto \left|\mu-\mu_c\right|^{-\nu}$. In this case, finite size effects should enter through the quantity $L/\xi$, where $L\equiv N^{1/d}$ with $d$ being the number of spatial dimensions. An equivalent scaling can also be written using $\left(\mu-\mu_c\right)L^{1/\nu}$; see Refs.~\cite{olsson2011,ClarkYielding} for further discussion on similar systems. Figure~\ref{fig:Unscaled}(d)-(f) shows that the data in Fig.~\ref{fig:Unscaled}(a)-(c) collapses according to
\begin{align}
F &= f\left(\left(\mu-\mu_c \right)L^{1/\nu_{\rm slip}}\right), \label{eq:F_scaling}
\\
\left\langle\gamma_{\rm slip}\right\rangle &= \left|\mu-\mu_c\right|^{-\beta_{\rm slip}}f_{\rm slip,\pm}\left(\frac{L}{\left|\mu-\mu_c\right|^{-\nu_{\rm slip}}}\right), \label{eq:slip_scaling}
\\
\left\langle\gamma_{\rm ms}\right\rangle &= \left|\mu-\mu_c\right|^{-\beta_{\rm ms}}f_{\rm ms,\pm}\left(\frac{L}{\left|\mu-\mu_c\right|^{-\nu_{\rm ms}}}\right). \label{eq:ms_scaling}
\end{align}
Here, $f_{\rm ms,\pm}$ and $f_{\rm slip,\pm}$ are dual-branch functions, with $+$ and $-$ denoting $\mu$ above or below $\mu_c$, respectively. Interestingly, we need two distinct values of $\nu$ for the initial strain, $\nu_{\rm ms}\approx 1.8$, and slip avalanche regime, $\nu_{\rm slip}\approx 1.1$ in 2D or $\nu_{\rm slip}\approx 0.8$ in 3D. The value $\nu_{\rm slip} \approx 1.8$ agrees with our previous result~\cite{ClarkYielding}, which was only calculated near to isostaticity; in Fig.~\ref{fig:Unscaled}, it is calculated far from isostaticity. The difference between $\nu_{\rm slip}$ and $\nu_{\rm ms}$ suggests that there are important differences in how MS states are accessed between these two regimes. We obtain the critical parameters $\mu_c$ and  $\nu_{\rm slip}$ by fitting the collapsed curves to appropriate functional forms using a Levenberg-Marquardt method~\cite{olsson2011,ClarkYielding}. We exclude system sizes with $N<N_{\rm min}$, and vary $N_{\rm min}$ until our fits become insensitive to our choice of $N_{\rm min}$. We also performed the corrections-to-scaling analysis in~\cite{VagbergFiniteSizeScaling}, which yields the same result we find with the scaling forms in Eqs.~\eqref{eq:F_scaling}-\eqref{eq:ms_scaling}. 

\begin{figure}[t]
\raggedright
(a) \hspace{40mm} (b) \\
\includegraphics[width=0.49\columnwidth]{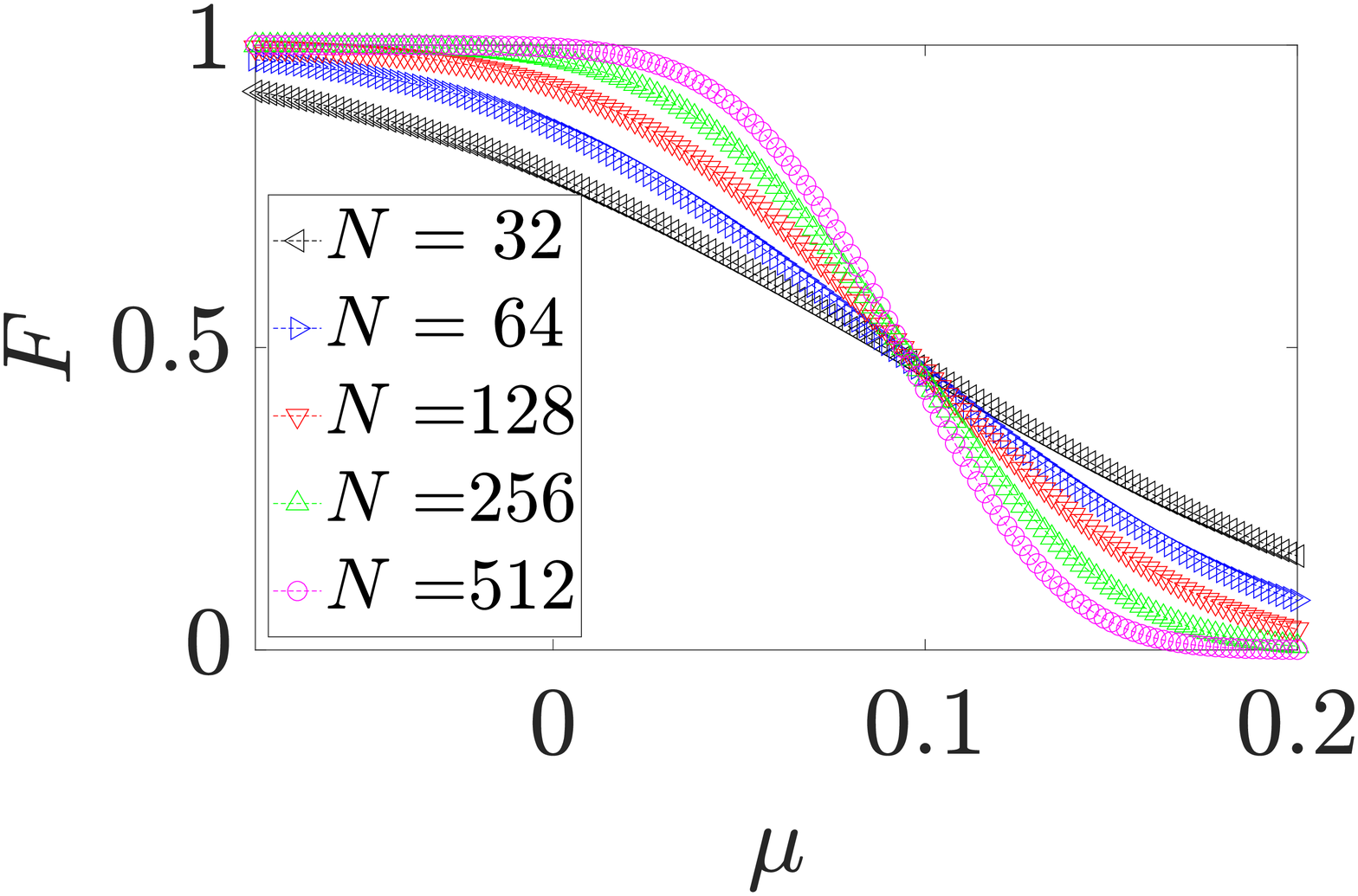}
\includegraphics[width=0.49\columnwidth]{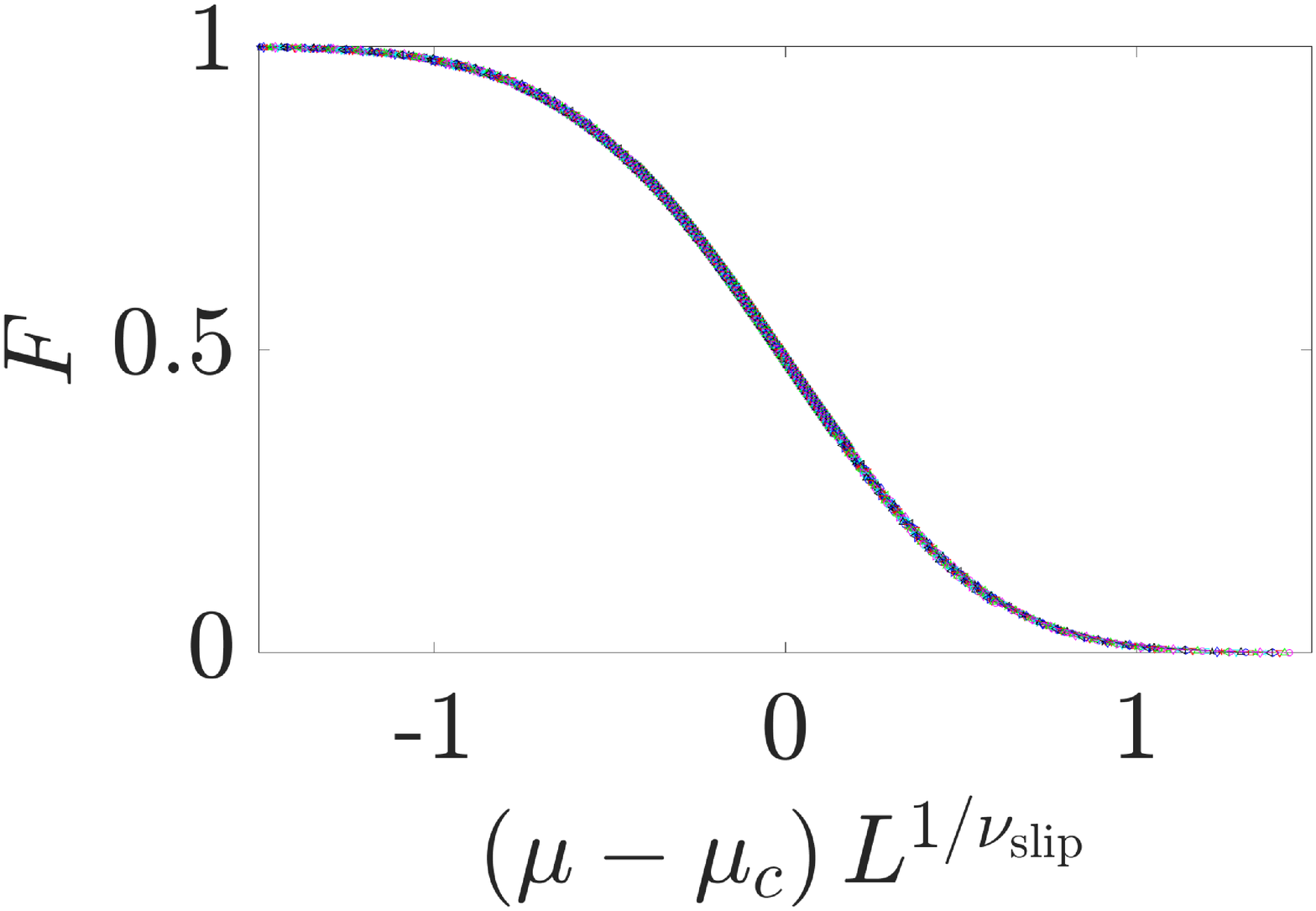}
\\
(c) \hspace{40mm} (d) \\
\includegraphics[width=0.49\columnwidth]{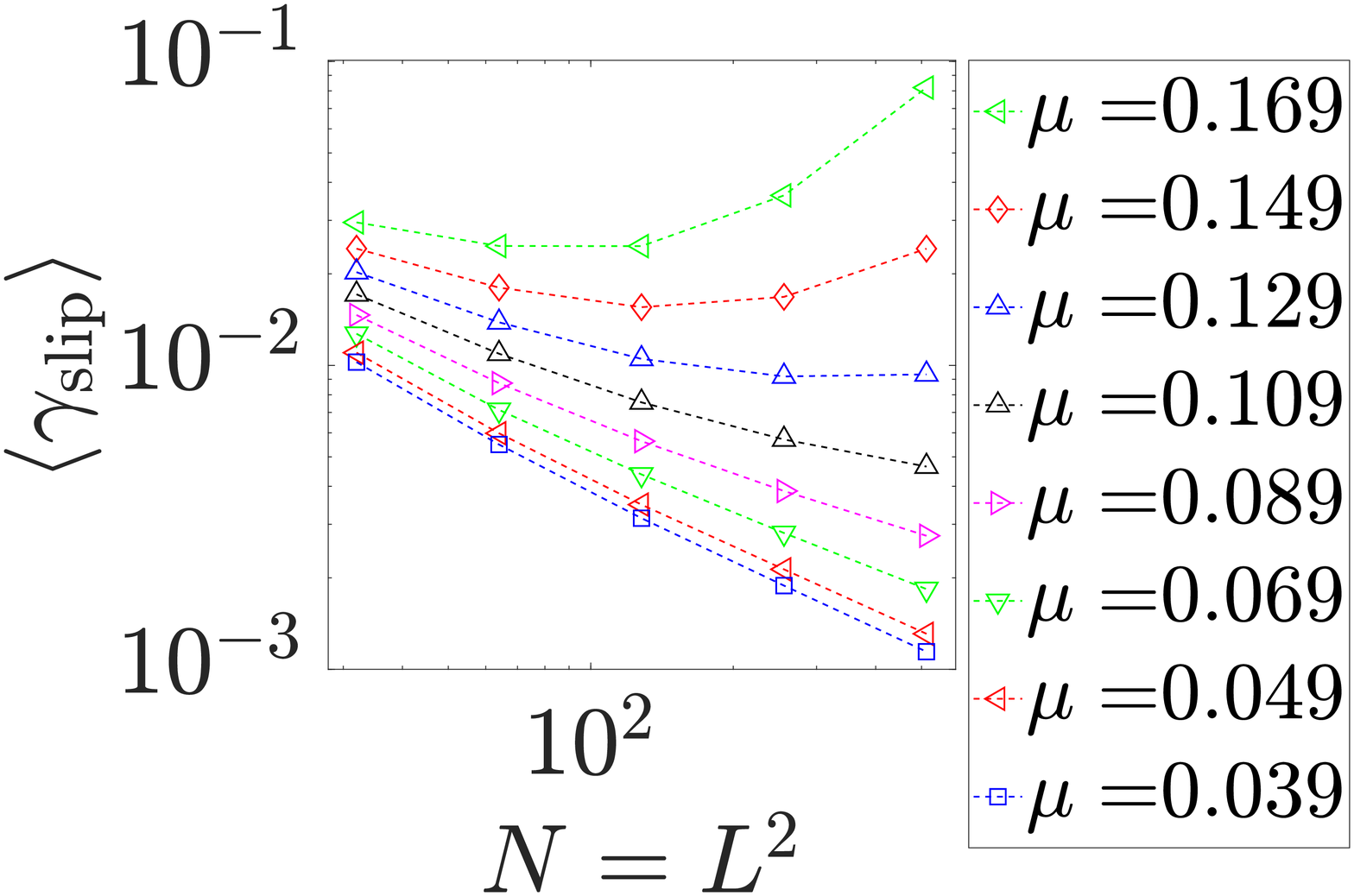}
\includegraphics[width=0.49\columnwidth]{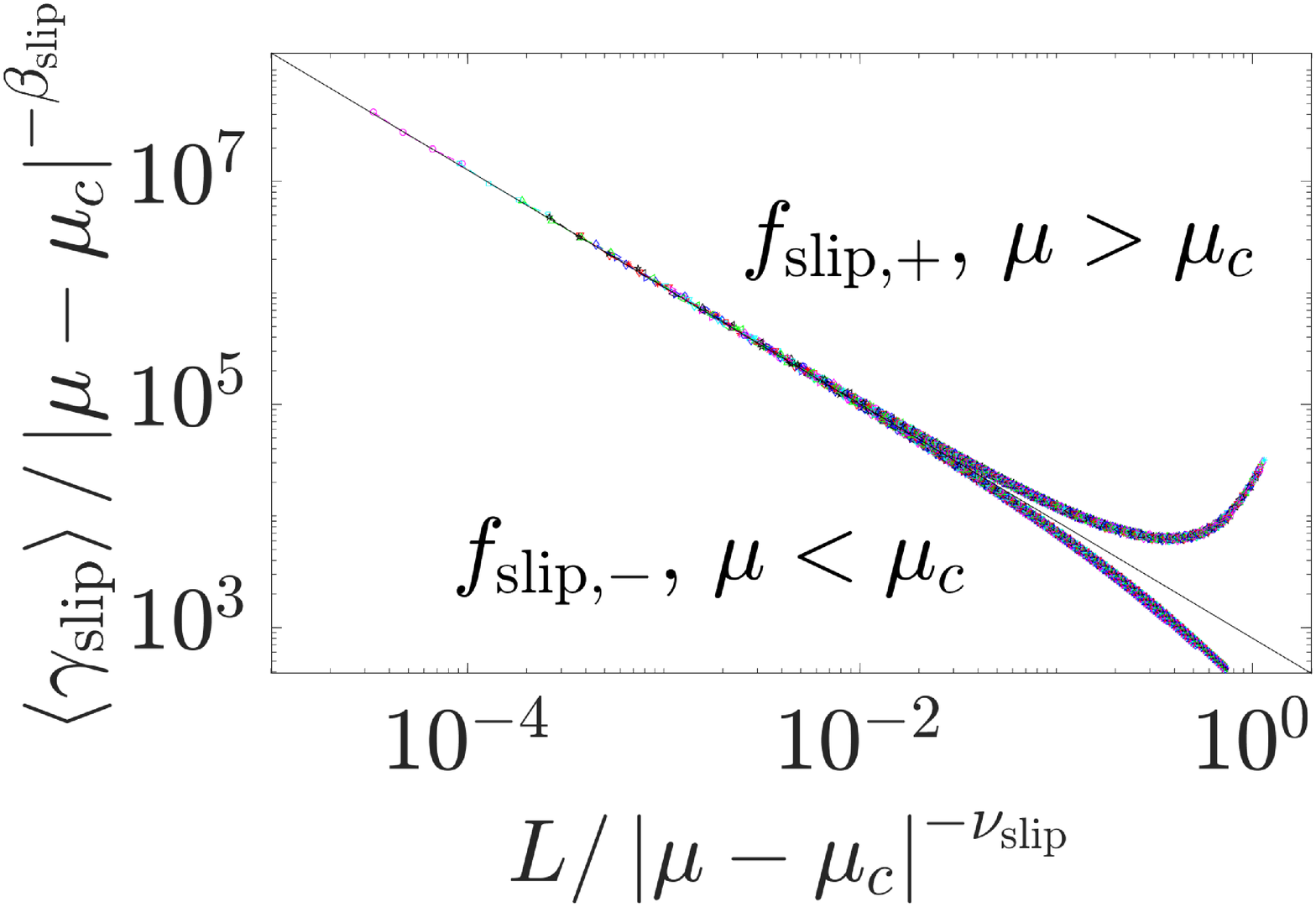}
\caption{Scaling collapses in 2D at low dimensionless pressure $\tilde{p}=0.0005$, $32\le N\le 512$. Fraction of states $F$ above $\mu$, unscaled (a) and scaled (b), with $\mu_c=0.097$ and $\nu_{\rm slip} = 1.10$. $\left\langle\gamma_{\rm slip}\right\rangle$ versus $N$, unscaled (c) and scaled (d), with $\mu_c = 0.097$, $\nu_{\rm slip} = 1.10$, and $\beta_{\rm slip}/\nu_{\rm slip} = -1.05$. In both cases, the unscaled plots show selected values of $N$ or $\mu$ while the scaled plots show all data.}
\label{fig:K2000}
\end{figure}

\begin{figure}
\raggedright
(a) \hspace{40mm} (b) \\
\includegraphics[width=0.49\columnwidth]{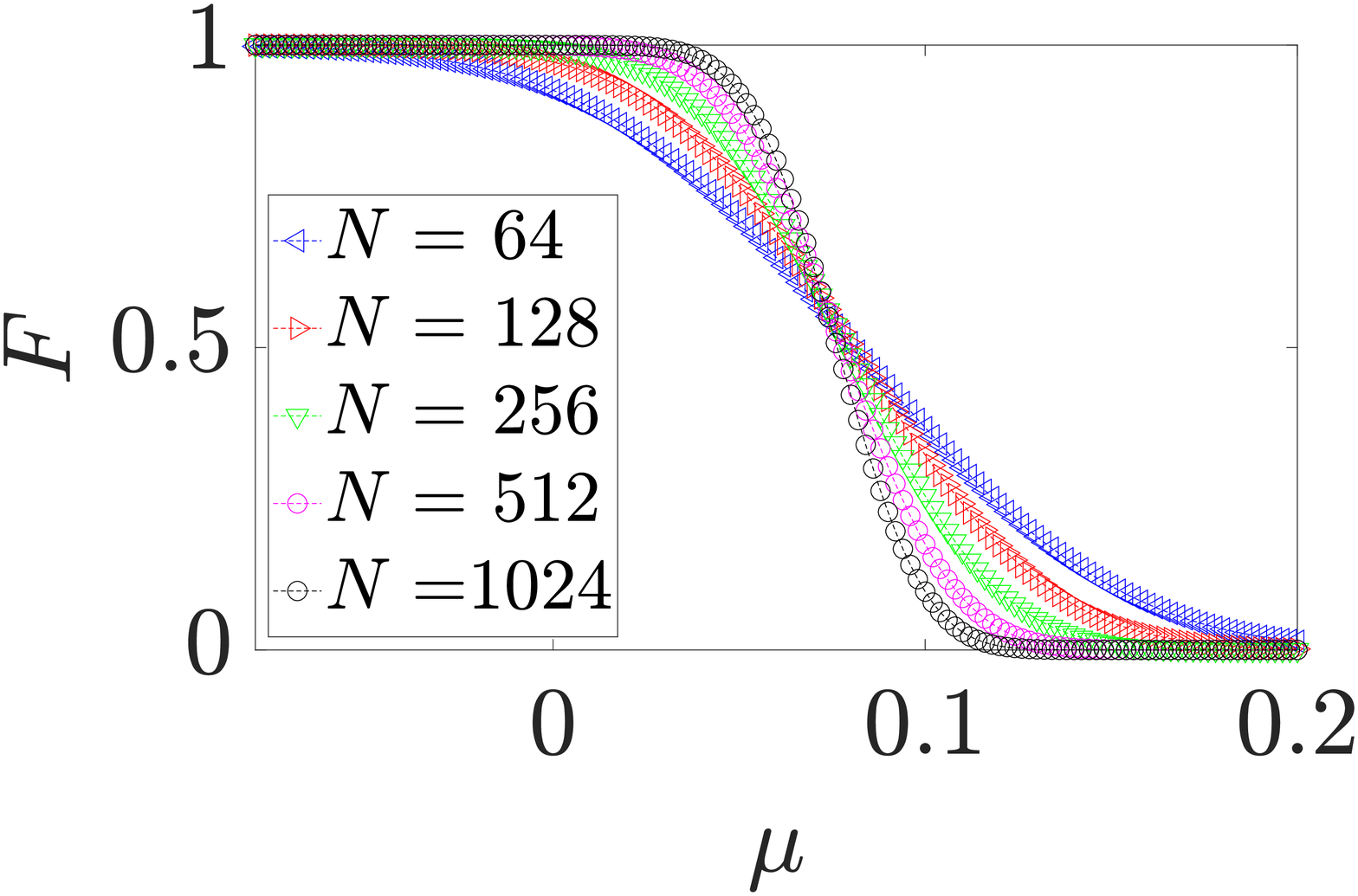}
\includegraphics[width=0.49\columnwidth]{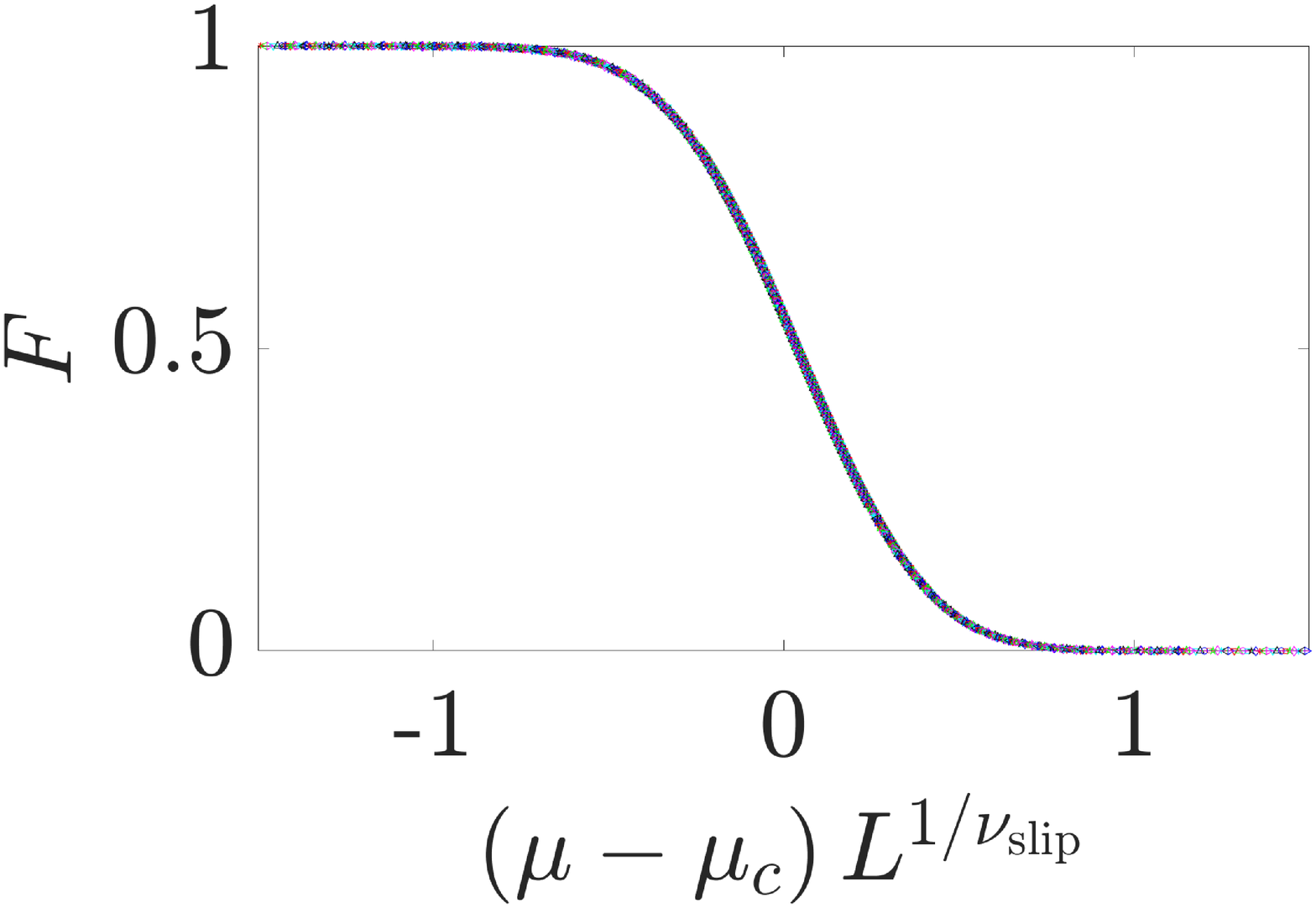}
\\
(c) \hspace{40mm} (d) \\
\includegraphics[width=0.49\columnwidth]{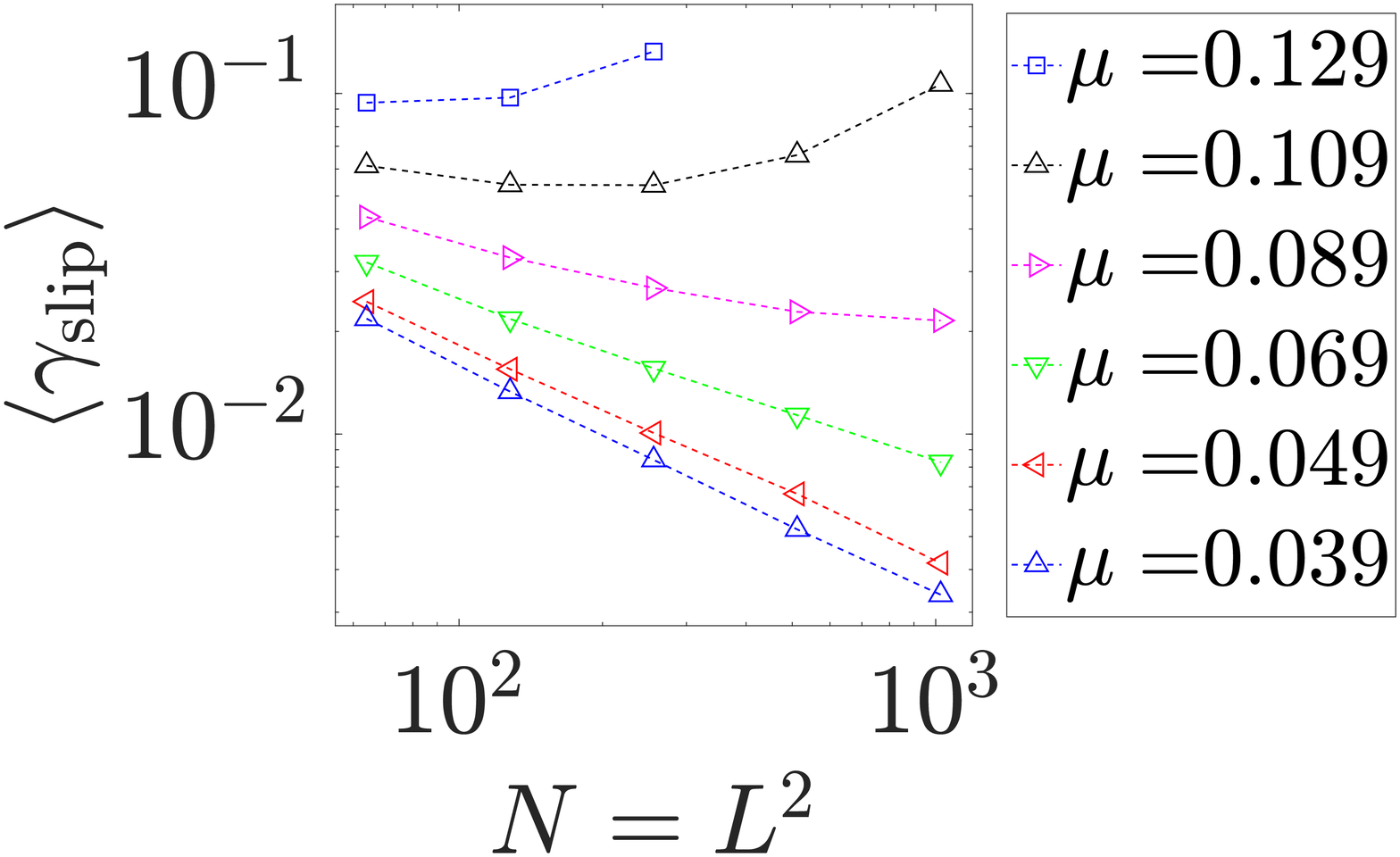}
\includegraphics[width=0.49\columnwidth]{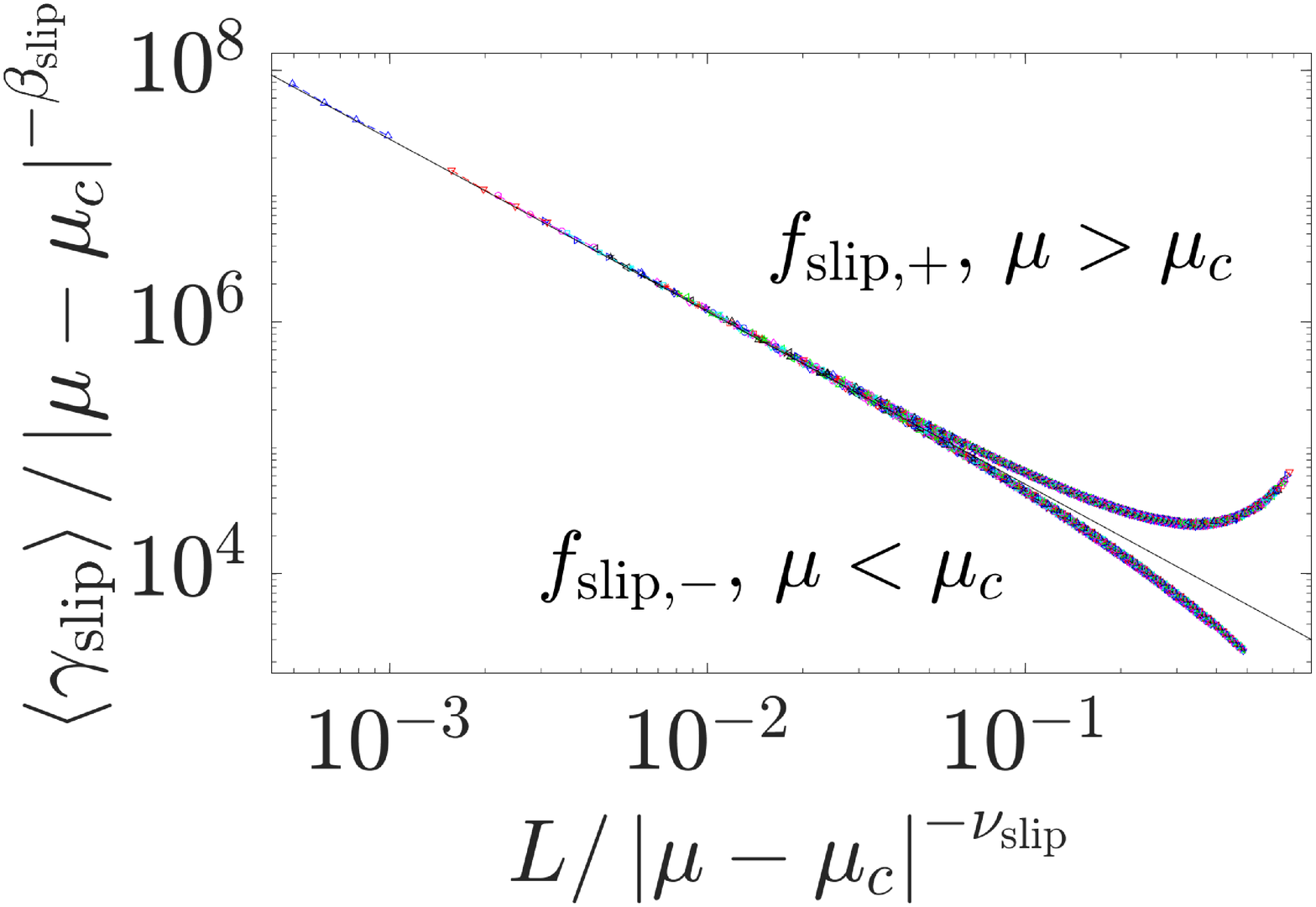}
\caption{Scaling collapses in 3D at high dimensionless pressure $\tilde{p}=0.05$, $64 \le N\le 1024$. Fraction of states $F$ above $\mu$, unscaled (a) and scaled (b), with $\mu_c=0.074$ and $\nu_{\rm slip} = 0.832$. $\left\langle\gamma_{\rm slip}\right\rangle$ versus $N$, unscaled (c) and scaled (d), with $\mu_c = 0.074$, $\nu_{\rm slip} = 0.832$, and $\beta_{\rm slip}/\nu_{\rm slip} = -1.37$. In both cases, the unscaled plots show selected values of $N$ or $\mu$ while the scaled plots show all data.}
\label{fig:3D}
\end{figure}

\begin{figure}
\raggedright
(a) \hspace{37mm} (b)\\
\includegraphics[width= 0.49\columnwidth]{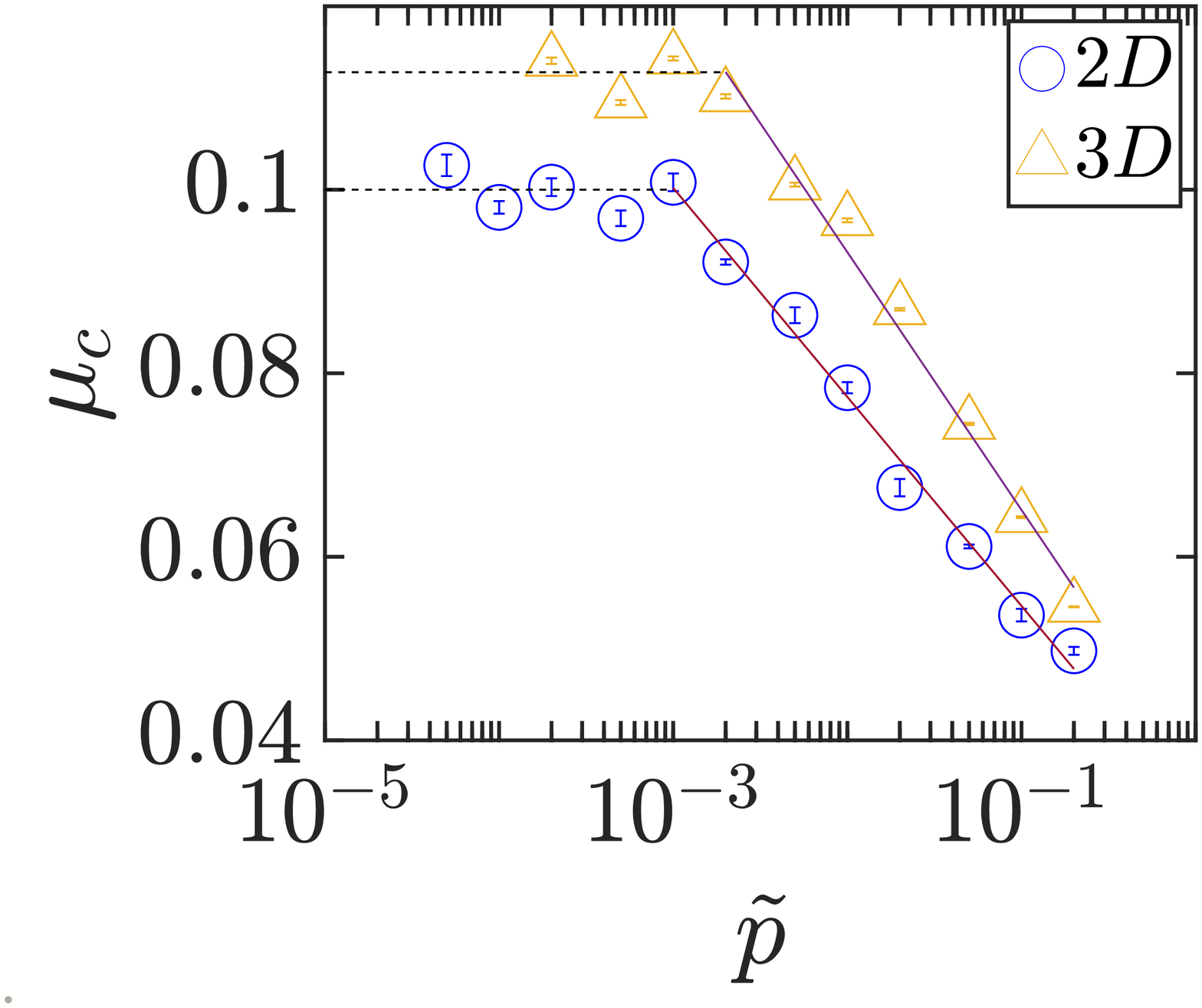}
\includegraphics[width=0.49\columnwidth]{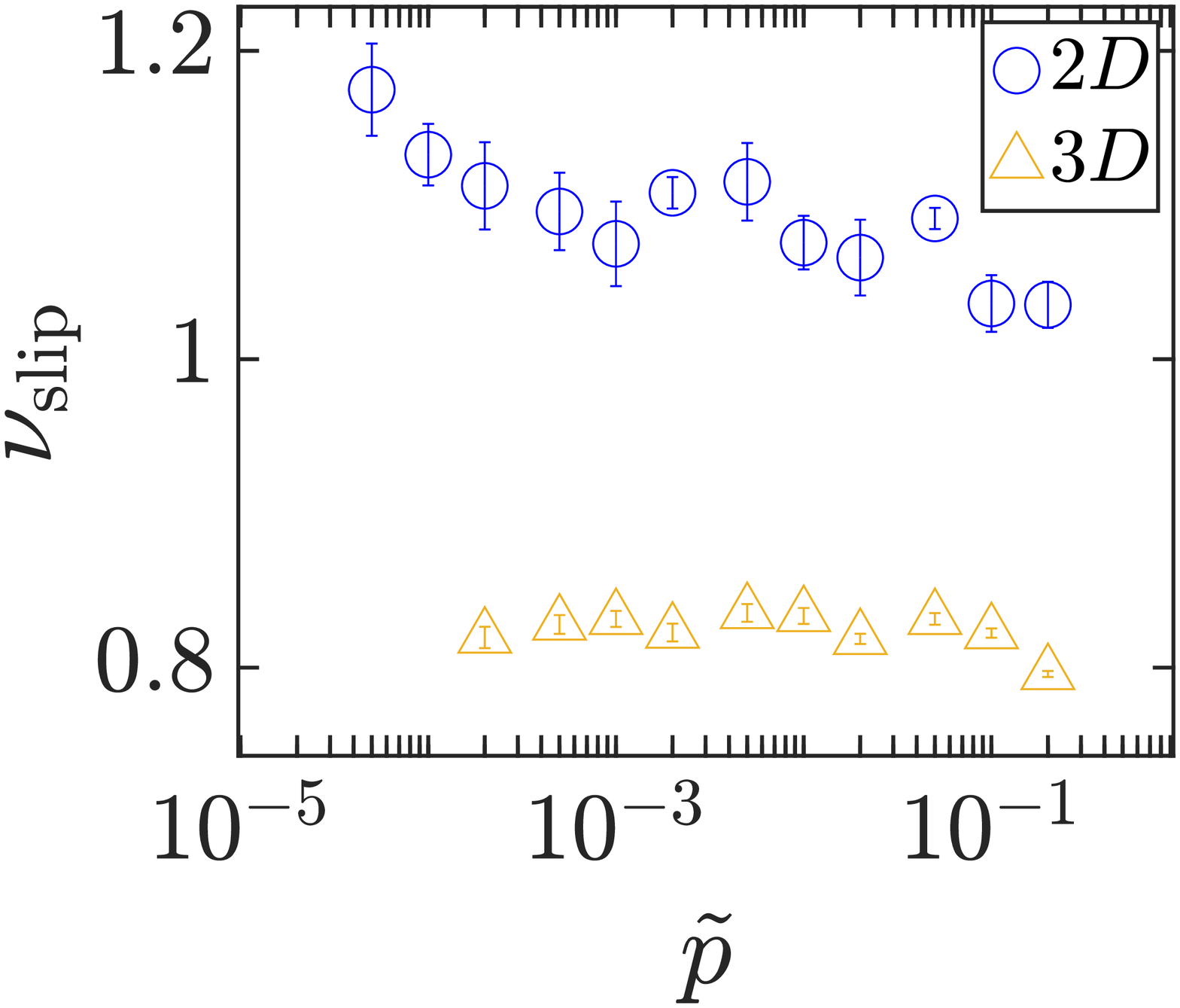}\\
(c)\\
\includegraphics[trim=0mm 0mm 20mm 0mm, clip, width=\columnwidth]{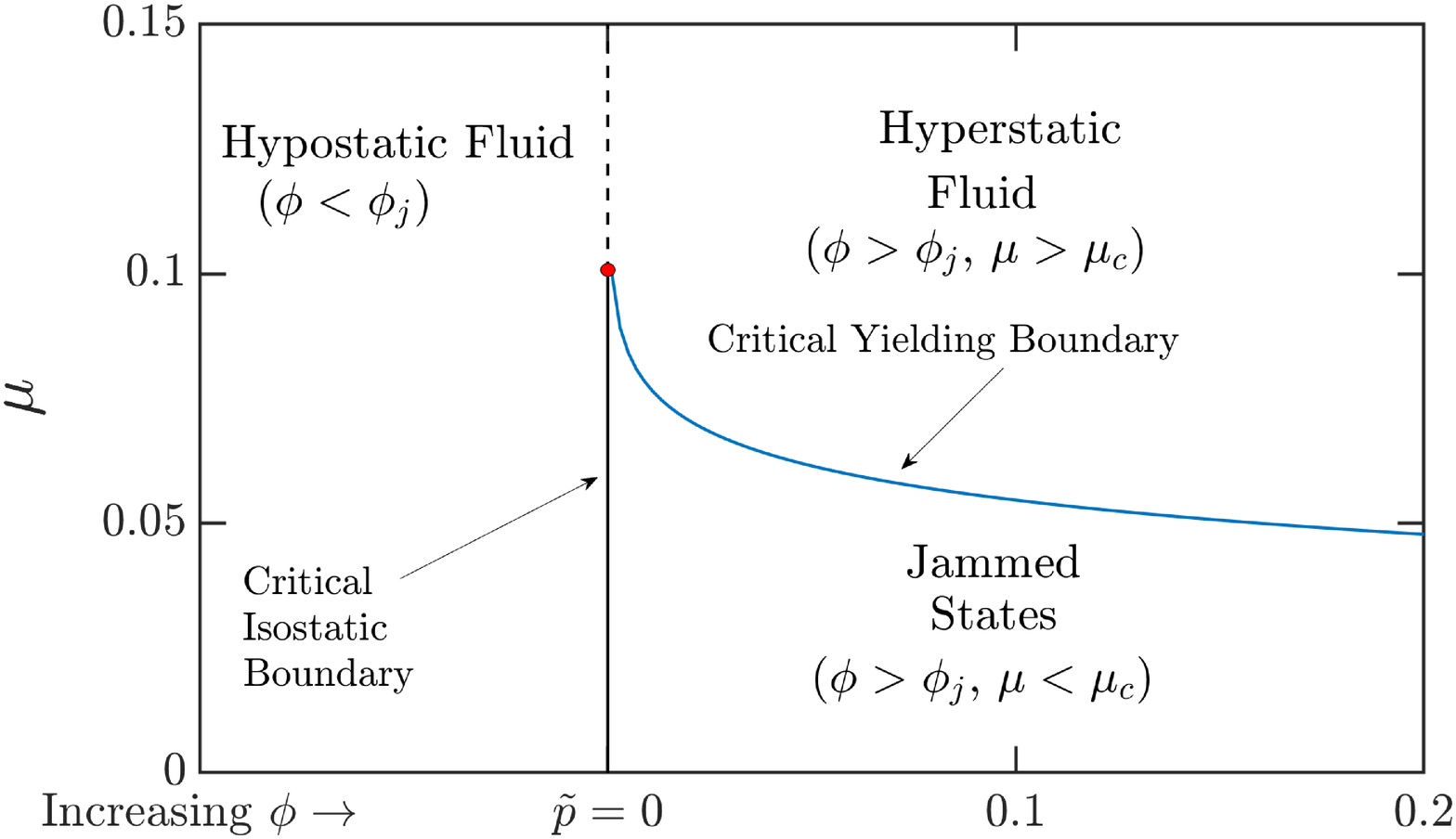}
\caption{(a,b) Values for (a) $\mu_c$ and (b) $\nu_{\rm slip}$ versus $\tilde{p}$, measured using the fitting protocol described in the text. Results in 2D and 3D are denoted by circles and triangles, respectively. Solid lines in (a) represent a linear fit of $\mu_c$ versus $\log \tilde{p}$ from $\tilde{p} = 10^{-3}$ through $2\cdot 10^{-1}$, while the dashed lines represent the large-stiffness limit. (c) Phase diagram summarizing our results. The solid blue line represents the estimation of $\mu_c(\tilde{p})$ from panel (a). 
}
\label{fig:Parameters}
\end{figure}

We then perform the same analysis for varying $\tilde{p}$ over the ranges $5\cdot 10^{-5} \le \tilde{p} \le 2\cdot 10^{-1}$ in 2D and $2\cdot 10^{-4} \le \tilde{p} \le 2\cdot 10^{-1}$ in 3D, spanning from near isostaticity (where $\xi_J$ is large) to far from isostaticity (where $\xi_J$ is small). The scaling description in Eqs.~\eqref{eq:F_scaling}-\eqref{eq:ms_scaling} and shown in Fig.~\ref{fig:Unscaled} holds for all values of $\tilde{p}$ in both 2D and 3D. We show data for an additional pressure in 2D, $\tilde{p} = 0.0005$, in Figure~\ref{fig:K2000}. We also show data in 3D with $\tilde{p} = 0.05$ in Fig.~\ref{fig:3D}. In both cases, the scaling functions are almost indistinguishable from those shown in Fig.~\ref{fig:Unscaled}. 
Figure~\ref{fig:Parameters} shows the critical parameters $\mu_c$ and $\nu_{\rm slip}$ plotted as a function of $\tilde{p}$. Each data point in Fig.~\ref{fig:Parameters} represents a fit of all data (as in Figs.~\ref{fig:Unscaled} through~\ref{fig:3D}) over many system sizes (typically $32 \le N \le 1024$) with 400 simulations per system size, so the plateau in Fig.~\ref{fig:Parameters} is not a system size effect. We find $\mu_c$ to be independent of $\tilde{p}$ for $\tilde{p} \le 10^{-3}$, and $\mu_c$ decreases logarithmically for $\tilde{p}>10^{-3}$, which agrees with~\citet{FavierDeCoulomb2017}. This occurs as excess contacts are added, which changes the structure of the force networks. 

However, we observe no similar crossover behavior as distance to isostaticity is varied in any other aspects of the scaling behavior. The critical exponents, as shown in Fig.~\ref{fig:Parameters}(b), and the scaling functions, as shown in Figures~\ref{fig:Unscaled},~\ref{fig:K2000}, and~\ref{fig:3D}, are highly insensitive to $\tilde{p}$, despite the wide variation in distance to isostaticity. Specifically, we find $\nu_{\rm slip}\approx 1.1\pm 0.1$ in 2D, $\nu_{\rm slip}\approx 0.8\pm 0.03$ in 3D, $\beta_{\rm slip}/\nu_{\rm slip} \approx -1 \pm 0.1$ in 2D, and $\beta_{\rm slip}/\nu_{\rm slip} \approx -1.3\pm 0.1$ in 3D. The uncertainty is estimated from the scatter in the data for different $\tilde{p}$, as seen in Fig.~\ref{fig:Parameters}. For the initial shear regime, we find that $\nu_{\rm ms}\approx 1.8$ is insensitive to $\tilde{p}$. However, $\beta_{\rm ms}/\nu_{\rm ms}$ appears to vary from roughly 0.2 at high $\tilde{p}$ to 0.6 at low $\tilde{p}$. This again points to potentially important differences between how MS states are explored between the slip avalanche and initial shear regimes and may have consequences for size-dependent arrest transitions~\cite{kamrin2015,ClarkYielding,Srivastava2018}.




\textit{Discussion.---} We have shown here that sheared amorphous soft sphere packings display finite size scaling that is consistent with a diverging length scale $\xi\propto|\mu-\mu_c|^{-\nu}$. The value of $\mu_c$ varies as $\tilde{p}$ is changed and extra contacts are added, but the forms of the scaling functions (as shown in Figures~\ref{fig:Unscaled} ,~\ref{fig:K2000}, and~\ref{fig:3D}) and the values of the critical exponents are nearly independent of distance to isostaticity over nearly four orders of magnitude in $\tilde{p}$. Considering the correlation length for isostaticity $\xi_J \propto |\phi - \phi_c|^{-\nu_J}$, if one assumes that $\nu_J$ is order unity~\cite{olsson2011} and $\tilde{p} \propto (\phi - \phi_c)$ for harmonic interactions~\cite{ohern2003}, then varying $\tilde{p}$ over this range represents $\xi_J$ varying over a similar range. This represents an enormous variation with respect to the isostatic critical point, implying that the distance to isostaticity does not control the critical behavior we demonstrate here. Our results suggest that yielding in, e.g., emulsions, foams, or granular materials is controlled by an underlying nonequilibrium critical transition that is distinct from isostaticity. We note that $\nu_{\rm slip}\approx 1.1$ for 2D and $\nu_{\rm slip}\approx 0.8$ in 3D are similar to the values $\nu\approx 1.1$ for 2D and $\nu\approx 0.7$ for 3D from Ref.~\cite{lin2014}, which presented a scaling description for yielding in amorphous materials~\cite{Miguel2001,Maloney2006,Salerno2012}. 

Figure~\ref{fig:Parameters}(c) shows the Liu-Nagel jamming phase diagram from, e.g., Refs.~\cite{ohern2003,Heussinger2009,bi2011} and many others, but with $\tilde{p}$ on the horizontal axis and $\mu=\tau/p$ on the vertical axis. The solid blue line represents the critical yielding boundary in 2D from Fig.~\ref{fig:Parameters}(a), and the solid vertical black line represents the isostatic critical transition. Jammed states exist only in the lower right region, above isostaticity and below the critical yielding boundary. Unjammed or fluid-like states can be either hypostatic ($Z<Z_{\rm iso}$ and $p=0$) or hyperstatic ($Z>Z_{\rm iso}$ and $p>0$). Some previous work on critical scaling near isostaticity has studied the onset of yield stress behavior under shear at varied $\phi$~\cite{olsson2007,Heussinger2009,Nordstrom2010,olsson2011,paredes2013rheology}. Such a system is situated at the ``triple point'' indicated by the red dot at the intersection of the jamming and yielding lines in Fig. \ref{fig:Parameters}(c). A complete theory may be able to unify these two critical transitions by a better understanding of the behavior at this point.

\bibliography{strengthening_refs2_Thompson}

\end{document}